\documentclass[aps,pre,showpacs,twocolumn]{revtex4}
\usepackage{epsfig}
\usepackage{amssymb}

\newcommand{\bi}{\begin{itemize}}
\newcommand{\ei}{\end{itemize}}

\newcommand{\be}{\begin{equation}}
\newcommand{\ee}{\end{equation}}
\newcommand{\bea}{\begin{eqnarray}}
\newcommand{\eea}{\end{eqnarray}}
\newcommand{\beastar}{\begin{eqnarray*}}
\newcommand{\eeastar}{\end{eqnarray*}}

\newcommand{\lav}{\left\langle}
\newcommand{\rav}{\right\rangle}
\newcommand{\eps}{\epsilon}
\newcommand{\rate}{\Gamma}
\newcommand{\ratee}{\tilde\Gamma}
\newcommand{\logtime}{\nu}
\newcommand{\dbar}{\bar{d}}
\newcommand{\xbar}{\bar{x}}

\newcommand{\half}{\frac{1}{2}}

\newcommand{\eq}[1]{~(\ref{#1})}

\newcommand{\order}{{{\mathcal O}}}

\newcommand{\ie}{{\it i.e.}}
\newcommand{\eg}{{\it e.g.}}
\newcommand{\ds}{\displaystyle}
\newcommand{\ddt}{\frac{\partial}{\partial t}}
\newcommand{\ddtau}{\frac{\partial}{\partial \tau}}
\newcommand{\p}{P}
\newcommand{\ptilde}{\tilde{P}}
\newcommand{\dd}{\tilde{d}}
\newcommand{\ddr}{\tilde{d}_{\rm r}}
\newcommand{\tr}{t_{\rm r}}
\newcommand{\dt}{\tilde{t}}
\newcommand{\dtau}{\tilde{\tau}}
\newcommand{\pup}{{\mathcal{P}}_1}
\newcommand{\pdown}{{\mathcal{P}}_0}
\newcommand{\psd}{{\mathcal{P}}_{\rm sd}}
\newcommand{\tst}{t_*}
\newcommand{\teq}{t_{\rm eq}}
\newcommand{\fpt}{\tau}
\newcommand{\mfpt}{\tau_{\rm mfp}}
\newcommand{\fptsc}{\tilde{\tau}}
\newcommand{\mfptsc}{\tilde\mfpt}
\newcommand{\up}{'}
\newcommand{\deq}{\bar{d}_{\rm eq}}
\newcommand{\scf}{f}
\newcommand{\del}{\delta}

\newcommand{\Ns}{N_{\rm s}}
\newcommand{\tw}{t_{\rm w}}

\begin{document}
\title{
Glassy dynamics in the East model
%Coarsening and equilibrium timescales in the East Model
%of glassy dynamics
}
\author{P. Sollich$^{1}$ and M. R. Evans$^{2}$}
\affiliation{
$^{1}$  Department of Mathematics,
        Kings College London, Strand, London WC2R 2LS, U.K.\\
$^{2}$  School of Physics, University of Edinburgh,
        Mayfield Road, Edinburgh EH9 3JZ, U.K.\\}
\date{\today}
\begin{abstract}
We study the dynamics of the East model, comprising a chain of
uncoupled spins in a downward-pointing field. Glassy effects arise at
low temperatures $T$ from the kinetic constraint that spins can only
flip if their left neighbour is up. We give details of our previous
solution of the non-equilibrium coarsening dynamics after a quench to
low $T$ (Phys.\ Rev.\ Lett.\ 83:3238, 1999), including the anomalous
coarsening of down-spin domains with typical size $\dbar \sim t^{T \ln
2}$, and the 
pronounced `fragile glass'-divergence of equilibration times as
$\tst=\exp(1/T^2\ln 2)$. We also link the model to the paste-all coarsening
model, defining a family of interpolating models that all have the
same scaling distribution of domain sizes. We then proceed to the
problem of equilibrium dynamics at low $T$. Based on a scaling
hypothesis for
the relation between timescales and lengthscales, we propose a model
for the dynamics of `superdomains' which are bounded by up-spins that
are frozen on long timescales. From this we deduce that the
equilibrium spin correlation and persistence functions should exhibit
identical scaling behaviour for low $T$, decaying as $g(\dt)$. The
scaling variable is $\dt=(t/\tst)^{T\ln 2}$, giving strongly stretched
behaviour for low $T$. The scaling function $g(\cdot)$ decays faster
than exponential, however, and in the limit $T\to 0$ at fixed $\dt$
reaches zero at a {\em finite} value of $\dt$.
\end{abstract}

\maketitle
%
% explanation of PACS numbers:
% ----------------------------
%
%
%
%%%%%%%%%%%%%%%%%%%%%%%%%%%%%%%%%%%%%%%%%%%%%%%%%%%%%%%%%%%%%%%%%%%%%%%%%%%%%%
\section{Introduction}
\label{Intro}
%%%%%%%%%%%%%%%%%%%%%%%%%%%%%%%%%%%%%%%%%%%%%%%%%%%%%%%%%%%%%%%%%%%%%%%%%%%%%%

The phenomenology of glassy systems---see
\eg~\cite{Jaeckle86,Angell95,EdiAngNag96,Debenedetti96} for excellent
reviews---has inspired many theoretical descriptions and explanations.
Experimentally, long relaxation times are observed; when these become
much longer than the observation timescale a glass transition is said
to occur.  Other signatures of glassy dynamics are correlation
functions that can be fitted by a stretched exponential decay law and
ageing phenomena~\cite{BouCugKurMez98} where, since the system is out
of thermal equilibrium, it keeps evolving as time goes by and
time-translation invariance is broken.

From a modelling perspective the same phenomenolgy arises when one
studies simple model systems by computer simulation.  Again,
relaxation times can outstrip the time available to run a simulation
and one never explores the equilibrium state.

The long relaxation times in glasses typically show a pronounced
divergence as the 
temperature $T$ is lowered and are often fitted experimentally by the
Vogel-Tammann-Fulcher (VTF) law
\begin{equation}
\tau = \tau_0\exp[-A/(T-T_0)]\;.
\label{VTF}
\end{equation}
The relaxation time $\tau$ may characterise, for example, the time for
a density fluctuation or an externally imposed stress to
relax. Although some heuristic justifications have been
offered~\cite{AdaGib65}, for practical purposes VTF is just a fit with
three parameters $\tau_0, A, T_0$. For $T_0=0$ it reduces to an
Arrhenius law.  A system for which $T_0$ is small, so that one has
something close to Arrhenius behaviour, is referred to as a `strong
glass', whereas a system exhibiting large deviations from Arrhenius
behaviour is called a `fragile glass'.  Generally, $T_0$ is much lower
than the experimental temperatures so that the mathematical
singularity in the fit (\ref{VTF}) is not physically relevant in an
experiment.  From the theoretical point of view, however, there has
been a long debate over whether $T_0$ might represent a true
thermodynamic transition temperature which would in principle be
measurable in the limit of infinitely slow cooling.

Although~(\ref{VTF}) is popular, it is not the only possibility for a
fit.  For example, the exponential inverse temperature squared (EITS)
form
\begin{equation}
\tau = \tau_0\exp(B/T^2)
\end{equation}
has been proposed as an alternative. This form does not exhibit a
singularity at any finite $T$.  Experimentally, or in a computer
simulation, it is difficult to distinguish between VTF and EITS
behaviour due to obvious limitations on the longest accessible
timescales; both can represent the experimentally observed $\tau(T)$
in many materials~\cite{RicBaes90}. Theoretical work is thus essential
for clarifying whether VTF or EITS might be more appropriate.
%
%This is the motivation
%for the work presented in this paper, where we study the low-$T$
%dynamics of a simple but nontrivial model with glassy behaviour.

Stretched exponential decay of a relaxation function, let us say an
autocorrelation $q(t)$, is expressed by the Kohlrausch-Williams-Watt
law
\begin{equation}
q(t) \sim \exp[- (t/\tau)^b]
\label{strexp}
\end{equation}
where the stretching exponent $b<1$.  An heuristic explanation
for this law is that there is a broad distribution $\Omega(\tau)$ of
relaxation modes with decay constants $\tau$,
\begin{equation}
q(t) = \int d\tau\ \Omega(\tau) \exp(-t/\tau).
\end{equation}
For example if one assumes $\Omega(\tau) \sim \exp (-a \tau)$ then for
large $t$ the dominant modes have $\tau=(t/a)^{1/2}$ which leads
to~(\ref{strexp}) with $b = 1/2$. This however leaves the physical
mechanisms by which such a relaxation time distribution would arise unclear.

One idea proposed to generate a broad distribution of relaxation times
from an explicit dynamical model was that of a hierarchy of degrees of
freedom~\cite{PalSteAbrAnd84}. The different levels in the hierarchy
then relax in series, the degrees of freedom in one level having to
wait for the degrees of freedom in the level below to reach some
configuration before they are free to evolve. This latter condition is
a realisation of a {\it kinetic constraint}.

A more concrete realisation of a kinetic constraint is the $f$-spin
facilitated kinetic Ising model introduced by Fredrickson and
Andersen~\cite{FreAnd84}.  Here no hierarchical structure needs to be
posited by hand to generate slow dynamics. All microscopic degrees
of freedom are of the same kind, namely non-interacting Ising spins on
a simple cubic
lattice in a downwards pointing field. However a spin can only flip
if at least $f$ nearest neighbour spins are pointing up (against the
field). For $f\geq 2$ this gives rise to slow cooperative relaxation,
as reviewed in \eg~\cite{RitSol03}. The physical motivation for the
kinetic constraint in this model becomes clear if one thinks of the
spins as coarse-grained density variables in a supercooled liquid,
with up and down-spins corresponding to low- and high-density
regions. The constraint then states
that a change in local density is only possible
if the overall density in the surrounding regions is low enough for
the particles to be able to rearrange.

An interesting modification of the Fredrickson-Andersen model arises
when the kinetic constraints are made anisotropic.
In the one-dimensional version considered
here, called the East Model~\cite{JaecEis91}, a spin can only flip if
its left (west) neighbour is pointing up, so that information
propagates only to the east. The rates for flipping mobile spins are 1
for down-flips and $\epsilon=\exp(-1/T)$ for up-flips, giving a small
equilibrium concentration $c=\eps/(1+\eps)$ of up-spins at low $T$
(see Sec.~\ref{Model} for details). The number of spins which are
mobile because their left neighbour is up will then also be small, and
this makes it plausible that the dynamics will slow down dramatically
as $T$ decreases.

In this paper we study the glassy dynamics
of the East model, both in equilibrium and out of equilibrium. We
focus on low temperatures $T$. This is the regime that is most
interesting since glassy features will be most pronounced; it is also
the regime where theoretical studies such as ours are most needed,
since the extremely long relaxation timescales make numerical
simulations difficult or impossible.

Before outlining the
structure of this paper, we give a brief review of existing work on
the East model. Much research to date has concerned the {\em
equilibrium} dynamics, as encoded \eg\ in relaxation functions such as
spin autocorrelation $C(t)$ or persistence $\pup(t)$ (see
Sec.~\ref{sec:functions} for definitions). Already when the model was
first proposed~\cite{JaecEis91} it was argued that relaxation
timescales should remain finite for any $T>0$, thus excluding a
transition to a nonergodic state where relaxation functions fail to
decay to zero. This has recently been proved rigorously: the longest
relaxation time, defined as the inverse of the smallest decay rate
that one would find by full diagonalization of the master equation, is
bounded between $\exp(1/2T^2\ln 2)$ and $\exp(1/T^2\ln 2)$ in the
limit of small temperatures~\cite{AldDia02}. The form of the bounds
demonstrates that the East model has EITS behaviour; we will find
below that it is the upper bound $\tst=\exp(1/T^2 \ln 2)$ that gives
the dominant low-$T$ behaviour of the relaxation times.

A number of approaches have been used to predict the actual shape of
the equilibrium relaxation functions. An `effective medium
approximation'~\cite{JaecEis91,EisJaec93} for $C(t)$ leads to a
self-consistency equation typical of mode-coupling approximations
(MCA) and predicts a spurious nonergodic transition at $c=0.5$
($\eps=1$); effectively the same result was later derived using
diagrammatic methods~\cite{PitAnd01}. The version of MCA derived by
Kawasaki~\cite{Kawasaki95} also gives a spurious transition, at
$c=0.2$ ($\eps=1/4$). Both approximations can therefore only be
reasonable at sufficiently large $\eps$, or for short times at smaller
$\eps$; a comparison with numerical simulations~\cite{PitYouAnd00}
shows that the effective medium approximation is generally more
accurate in these regimes. Improved diagrammatic
resummations~\cite{PitAnd01} avoid the prediction of a spurious
transition, and are quantitatively more satisfactory over a larger
range of $t$ and $\eps$. However, for small $\eps$ they still predict
a decay of $C(t)$ that is too fast and too similar to an exponential
compared with numerical simulations. Similar comments apply to the
results of~\cite{MauJaec99}, where an adiabatic approximation together
with the hierarchical nature of the relaxation processes---discussed
in detail below---was used to study the analogue of
$C(t)$ for finite chains. The 
non-exponential behaviour of $C(t)$ had been noticed early
on~\cite{JaecEis91}, but is well fitted by a stretched exponential
only over a limited time range. We conjectured in~\cite{SolEva99}, and
will find below, that the correct scaling variable for $C(t)$ at low
$T$ is indeed of a `stretched' form, $\dt=(t/\tst)^{T \ln 2}$, but that
the relevant scaling function decays more quickly than an
exponential. The stretching exponent $T\ln 2$ was also found to govern
the decay of the up-spin persistence
$\pup(t)$~\cite{BuhGar01,BuhGar02}.

Recently, interest in the East model has shifted to more complicated
features of the equilibrium dynamics, \eg\ the existence of dynamical
heterogeneities~\cite{GarCha02}, and to the out-of-equilibrium
behaviour. In~\cite{MunGabInaPie98}, nonlinear relaxation processes
after large changes in $T$ were simulated. Otherwise, interest has
centred on the behaviour after a quench from high to low temperature.
For a quench to exactly $T=0$, the dynamics is exactly
solvable~\cite{CriRitRocSel00,DesGodLuc02}. The solution is
essentially equivalent to that of the corresponding model with
isotropic constraints, \ie\ the one-dimensional Fredrickson-Andersen
model with $f=1$~\cite{FolRit96,SchTri97}, and in fact also applies to
a whole family of models that interpolate between the isotropic and
anisotropic limits~\cite{BuhGar01,BuhGar02}. For a quench to nonzero
$T$, the autocorrelation function $C(t,\tw)$ between spins at times
$\tw$ and $t$ after the quench
and the corresponding response were simulated
in~\cite{CriRitRocSel00}. The correlation $C(t,\tw)$ exhibits
plateaux, which we will see in Sec.~\ref{sec:solution} can be
rationalized from the hierarchical nature of the dynamics.
Non-equilibrium steady states caused by applying an external `drive'
have recently also been studied, using a `tapping dynamics' inspired
by ideas from granular media~\cite{BerFraSel02} as well as
`rheological driving' designed to model the effect of a shear
flow~\cite{Fielding02}. Finally, we mention an interesting
two-dimensional spin model, which can be mapped onto a system of
non-interacting defects with kinetic constraints and turns out to have
low-$T$ behaviour very similar to that of the East
model~\cite{NewMoo99,GarNew00,Garrahan02}.

The paper is organised as follows. After defining the East model fully
in Sec.~\ref{Model}, we turn to the out-of-equilibrium dynamics in
Sec.~\ref{sec:noneq}, giving details of the results announced
in~\cite{SolEva99}. After a quench to low temperature, the equilibrium
concentration of up-spins at the new $T$ is small compared to its
initial value. Thus up-spins are eliminated, essentially irreversibly,
and the dynamics can be viewed as a coarsening process whereby
down-spin domains coalesce as up-spins disappear. We find the scaling
of the rates at which domains of length $d$ disappear, which follow a
hierarchical pattern so that the coarsening dynamics splits into
well-separated stages. This allows us to find an exact solution for
$T\to 0$. The overall scaling of the rates, as $\rate(d) \sim
d^{-1/T\ln 2}$, also implies that the system exhibits anomalous
coarsening, with typical domain sizes increasing with time as $\bar{d}
\sim t^{T\ln 2}$. Extrapolating to equilibrium domain lengths
$\deq\approx 1/\eps$ (see below) then gives the dominant divergence of
the equilibration time as $\tst \sim \exp(1/T^2\ln2)$. Interestingly,
in the scaling limit of the out-of-equilibrium dynamics ($1\ll
\bar{d}\ll \deq$), we find that the domain size distribution is
identical to that of the paste-all model~\cite{DerGodYek91}. We
rationalize this observation, and provide in
Sec.~\ref{sec:paste-all} a new family of models with identical scaling
distributions that interpolate between the paste-all and East model
limits.

In Sec.~\ref{sec:eq} we shall then explore the {\em equilibrium}
dynamics. Based on a plausible conjecture regarding the scaling of
relaxation times with distance at low $T$, we introduce a simplified
picture which we refer to as the {\em superdomain model}. This should
capture the essence of the equilibrium dynamics and become exact in
the $T\to 0$ limit.  The model enables us to probe the long timescales
of the equilibrium dynamics numerically, and leads us to a new scaling
form for the equilibrium spin autocorrelation function, $C(t)=g(\dt)$,
with $\dt=(t/\tst)^{T \ln 2}$. This shows strong stretching for $T\to
0$, but it will turn out that the scaling function $g$ decays more
quickly than an exponential, actually decreasing to zero at a finite
value of $\dt$. Finally, a brief summary of our results and outlook to
future work is given in Sec.~\ref{sec:conclusion}.

\section{Model definition}
\label{Model}

The model comprises $L$ Ising spins  $s_i=0,1$ on a one-dimensional lattice 
with periodic boundary conditions (site $i=L+1$ is
identified with site $i=1$). The dynamics are defined by the following
spin-flip rates
\begin{equation}
\begin{array}{ccl}
1\ 1 \to 1\ 0 & \mbox{with rate} & 1\\[1ex]
1\ 0 \to 1\ 1 & \mbox{with rate} & \ds \epsilon =\exp(-1/T)\;.
\end{array}\label{dynamics}
\end{equation}
Thus a spin can only flip if its left neighbour is pointing up (note
that in the original paper~\cite{JaecEis91} the mirror image of the
above definition was used, so that the {\em right} neighbour had
to point up for a spin to be able to flip). By a rate, say $x$, we
mean that in a small time $dt$ the event happens with probability $x\,
dt$. It is easy to check that the dynamics obey detailed balance with
respect to the energy function $E = \sum_{i=1}^L s_i$, \ie\ the
equilibrium distribution corresponds to free spins in a downwards
pointing field:
\begin{equation}
P_{\rm eq}(\{s_i\}) = \frac{1}{Z} \exp \left( -\frac{1}{T}\sum_i s_i \right)
= \frac{\epsilon^{\sum_i s_i}}{(1+\epsilon)^L}\;.
\label{Peq}
\end{equation}
It follows that the equilibrium concentration of up-spins, $c =
\langle s_i \rangle$, is given by
\begin{equation}
c = \frac{\epsilon}{1+\epsilon}\;.
\label{c}
\end{equation}
To show that this is the unique steady state we require that the
(finite) system is ergodic, \ie\ that any configuration can be
reached from any other. This is true for all configurations except for
the one with all spins down; this is a configuration which can be neither
entered nor left and which we therefore ignore. To see that the rest of the
configuration space is ergodic, note that from any configuration with
a non-zero number of up-spins one can flip spins up until the
`all up' configuration is reached. Then, to obtain any desired
configuration one flips spins down in an ordered way to create the
appropriate regions of down-spins.

The basic objects that we use for the description of the system are
{\em domains}. As shown by the vertical lines in
% the sample configuration
$$
\ldots0|1000|1|1|10|100|1|1|10|10\ldots,
$$
a domain consists of an up-spin and all the down-spins that separate
it from the nearest up-spin to the right. (This convention is opposite
to that of~\cite{SolEva99}, but leads to equivalent conclusions and
will be more convenient in our treatment of equilibrium dynamics in
Sec.~\ref{sec:eq} below.) The length $d$ of a domain then gives the
distance between the up-spin at its left edge and the next up-spin to
the right. Note that adjacent up-spins are counted as separate domains
of length $d=1$. In equilibrium, it follows from~(\ref{Peq}) that the
domain lengths are geometrically distributed
\begin{equation}
P_{\rm eq}(d)=\epsilon/(1+\epsilon)^d
\label{pd_equil}
\end{equation}
with mean
\begin{equation}
\deq=\frac{1+\eps}{\eps}\; .
\label{deq}
\end{equation}
As explained, our aim will be to
pursue analytical calculations where possible, but we explain briefly
how simulations were carried out. We used a continuous-time
BKL~\cite{BorKalLeb75} algorithm, where the time intervals between
spin flips are sampled directly; a standard Monte Carlo algorithm
where spin flips are first proposed and then accepted or rejected
would be much slower at low $T$. In the BKL algorithm, once the time
interval to the next flip is determined one decides probabilistically
which of the mobile spins to flip, choosing each with a probability
$p_i$ proportional to its flip rate (1 or $\eps$). A simple method for
doing this would be to sample a uniform random variable $r$ on $[0,1]$
and then go through the mobile spins until the spin $i$ is found for
which $\sum_{j=1}^{i-1} p_j < r \leq \sum_{j=1}^{i} p_j$. This search
for $i$ takes $\order(L)$ steps, however, and in fact for large $L$
quickly becomes the most computationally intensive part of the
algorithm. Instead, we define partial sums $q_{j,a}$ over all blocks
of spins of length $2^a$, where $a=0,\ldots, l$ if the total chain
length is $L=2^l$. The $q_{j,a}$ can be defined recursively by
$q_{j,a=0}=p_j$ and
\[
q_{j,a} = q_{2j-1,a-1}+q_{2j,a-1} \qquad (j=1,\ldots, L/2^a)
\]
(In fact, we work with analogous partial sums for the number of mobile
up- and down-spins separately, from which the $q_{j,a}$ can easily be
retrieved; this allow us to use fast integer arithmetic.) The
$q_{j,a}$ can be thought of as arranged on a binary tree, and finding
the spin $i$ to flip becomes a simple walk from the top level $a=l$ to
the bottom level $a=0$, branching left or right on each level to keep
$r$ within the $q_{j,a}$-values bounding the remaining subpart of the
tree. This takes $l\sim\ln L$ steps,
and the $q_{j,a}$
can be updated in similar time once a spin has been flipped, giving a
reduction in computing time $\sim (\ln L)/L$. Compared to earlier
simulations, \eg~\cite{MunGabInaPie98}, rather longer timescales (up
to $t=10^{11}$) can be accessed with this method.

\section{Non-equilibrium dynamics}
\label{sec:noneq}

\subsection{Coarsening dynamics}

In this section we consider the dynamics of the East model after a
quench from equilibrium at some high initial temperature $T \gg 1$ to
low temperature $T\ll 1$ ($\epsilon \to 0$)~\cite{SolEva99}. At the new
temperature the equilibrium concentration of up-spins is much smaller
than before, so that the number of up-spins must decrease in time.
Correspondingly, the typical domain sizes must grow: the system coarsens.

To understand the nature of this coarsening process, recall first that
the equilibrium concentration of up-spins is, from~(\ref{c}),
$c=1/d_{\rm eq}=\epsilon+\order(\epsilon^2)$. Hence the equilibrium
probability of finding an up-spin within a chain segment of {\em
finite} length $d$ is $\order(d\epsilon)$ and tends to zero for
$\epsilon\to 0$. In the limit
\begin{equation}
\epsilon\to 0\quad\mbox{at fixed}\quad d\;,
\label{limit}
\end{equation}
the flipping down of up-spins therefore becomes {\em irreversible to
leading order}~\cite{irrev}.
In terms of domains, this means that the coarsening
dynamics of the system is one of coalescence of domains: an up-spin
that flips down merges two neighbouring domains into one large domain.
Note that the limit~(\ref{limit}) does not extend to include
equilibrium domain lengths that scale as $d\sim 1/\eps$. Therefore the
irreversible coarsening is a purely non-equilibrium phenomenon and
does not pertain to the equilibrium dynamics; this is why at low $T$
the former is in fact {\em easier} to analyse than the latter, as we
shall see.

Irreversible coarsening processes such as the one above have been
studied in a variety of contexts.  In particular, if the rate of
elimination depends solely on the domain size and not on the sizes of
neighbouring domains there is a very convenient property: during such
a process, no correlations between the lengths of neighbouring domains
can build up if there are none in the initial state.  The proof is an
easy generalisation of that of~\cite{BraDerGod94}. We include it for
completeness here. Consider first an initial arrangement of $N$
domains on a ring. There are $N!$ ways for this system to coarsen, \ie\
$N$ possibilities for the first domain to disappear, $N-1$
possibilities for the second etc. Each possibility is weighted with a
probability which is a function of the rates of disappearance of the
$N$ domains.  Summing over the $(N-1)!$ possible initial arrangements
of the $N$ domains on the ring implies $N!(N-1)!$ possible `histories'
for the dynamics.

Now note that if the domains are uncorrelated then the dynamics
becomes equivalent to a mean-field model (often referred to as a bag
model or an independent intervals theory): when a domain is eliminated
one picks a domain at random to be its right neighbour and coalesces
the two.  In this dynamics there are at each step $N(N-1)$
possibilities: the factor $N$ comes from which domain is eliminated
and the factor $N-1$ from which domain is picked to be its neighbour.
Therefore for $N$ initial domains there are $N!(N-1)!$ possible
histories, each weighted by the probability for the domains to be
eliminated in the specified sequence.  These possible histories are in
one-to-one correspondence with the possible histories for the full
model when summed over all possible initial arrangements. One also
easily verifies that the probabilities with which these histories
occur, as well as the distributions for the times that separate
successive events within each history, are the same in both
cases. This proves that throughout the irreversible
coarsening process the domains are uncorrelated.

\subsection{Energy barriers}
\label{sec:barriers}

We now estimate the typical timescale $\Gamma^{-1}(d)$ for the
disappearance of domains of length $d$ through coalescence with their
right neighbours.  This then defines a typical rate $\Gamma(d)$ for
the elimination of domains.  Because domain coalescence corresponds to
the flipping down of up-spins, $\Gamma(d)$ can also be defined as
follows. Consider an open spin chain of length $d$, with a `clamped'
up-spin ($s_0=1$) added on the left.  Starting from the state
$(s_0,s_1, \ldots, s_d)$ = $10\ldots 01$, $\Gamma^{-1}(d)$ is the
typical time needed to reach the `empty' state $10\ldots 00$ where spin
$s_d$ has relaxed, \ie\ has flipped down.  Any instance of this
relaxation process can be thought of as a path connecting the two
states. Let us call the maximum number of `excited' spins (up-spins
except $s_0$) encountered along a path its height $h$. One might think
that the relaxation of spin $s_d$ needs to proceed via the state
11\ldots 1, giving a path of height $d$. In fact, the minimal path
height $h(d)$ is much lower and given by
\be
h(d)=n+1 \quad \mbox{for}\quad 2^{n-1} < d \leq 2^n
\label{hierarchy}
\ee
where $n=0, 1, \ldots$

\begin{figure}
\[
\begin{array}{llc}
d=1\quad
&
1 \,\underline{1}  \to \,1\,0
&\quad h(1) = 1\\[1ex]
d=2\quad
&
1 \,0\, 1  \to \,1\underline{\,1 \,1} \to \,1\,1\,0 \to \,1\,0\,0
&\quad h(2) =2\\[1ex]
d=3\quad
&
1 \,0\,0\, 1  \to \,1\,1\,0 \,1 \to \,1\,\underline{1\,1\,1}\to&\\[1ex]
& \cdots \to \,1\,0\,0\,0 &\quad h(3) =3\\[1ex]
d=4\quad
&
1 \,0\,0\,0\, 1 \to  \cdots \to 1\,\underline{1\,1}\,0 \,\underline{1} \to
&\\[1ex]
&
%\hspace{1.5cm} \,
1\,0\,1\,0 \,1 \to 1\,0\,\underline{1\,1\,1} \to 
\cdots \to &\\[1ex]
& 1\,0\,1\,0\,0 
\to \cdots \to 1
\,0\,0\,0\,0 &\quad h(4) =3
\end{array}
\]
\vspace*{0.2cm}
\caption{Elimination of a domain of size $d$. Shown are paths through
spin configurations that traverse the minimum energy barrier. The
height of the barrier is $h(d)$ and the excited spins are underlined
in the highest energy configuration(s) along the path.
\label{fig:barrier}
}
\end{figure}
To get a feeling for the result (\ref{hierarchy}) consider in
Fig.~\ref{fig:barrier} some small domain sizes.
The figure illustrates that to generate an up-spin adjacent to the
one to be relaxed (which is the first spin of the next domain on the
right) one can proceed via a sequence of
stepping-stones, \eg\ for $d=4$ one first generates an isolated up-spin
in the middle of the domain then uses this stepping-stone to generate
the subsequent excited spins in a similar manner to the relaxation of
a $d=2$ domain. Iteration of this argument straightforwardly
proves\eq{hierarchy} for $d=2^n$~\cite{MauJaec99}.

In order to prove~(\ref{hierarchy}) generally we introduce the
quantity $l(h,k)$ which is defined for the open chain discussed
above. It is the length of the longest configuration that contains
exactly $k$ up-spins, always counted ignoring the fixed up-spin $s_0$,
and that can be reached from an initially empty chain (again, except
for $s_0$) along some path of height $h$.  The length of a
configuration is defined here as the position of the furthest up-spin
to the right.  Note that $l(h,k) > l(h,k-1)$; thus $l(h,h)$ gives the
longest configuration that can be reached along any path of height
$h$.

Now consider how a configuration realising $l(h,k)$ could be
constructed.  First one generates an isolated up-spin as far to the
right as possible. This distance is $l(h,1)$. Then one starts from the
up-spin just generated to generate a second up-spin as far to the
right of the first as possible.  Since the first up-spin remains up,
there is one less unit of energy to play with to generate the second
up-spin and the total distance from the origin will be $l(h,1) +
l(h-1,1)$.  Continuing in this fashion to generate all $k$ up-spins,
one arrives at
\begin{equation}
l(h,k) = \sum_{m=1}^k l(h-m+1,1)\;.
\label{recur1}
\end{equation}
To close the set of equations (\ref{recur1}) we need an expression for
$l(h,1)$. This can be obtained by invoking the reversibility of paths:
since the dynamics obeys detailed balance, the existence of a path of
height $h$ from an initial to a final state implies that the reversed
path connects the final to the initial state and is also of height $h$.
Thus $l(h,1)$ gives the size of the longest domain that can be relaxed
along some path of height $h$. It is easy to see that the final step
in this process requires an up-spin adjacent to the up-spin which is
to be relaxed. Since, with this latter up-spin, we have already used up
one unit of energy or height, the maximum distance from the origin at
which such an up-spin could be generated is $l(h-1,h-1)$. Therefore we
conclude
\begin{equation}
l(h,1) = l(h-1,h-1)+1\;
\label{recur2}
\end{equation}
with boundary condition $l(0,0)=0$.
Inserting~(\ref{recur2}) into~(\ref{recur1}) gives
\begin{equation}
l(h,k) = \sum_{m=1}^{k} l(h-m,h-m) +k \;.
\label{recur3}
\end{equation}
Setting $k=h$, (\ref{recur3}) has solution
$l(h,h) = 2^h{-}1$. Substituting this back into~(\ref{recur3}) yields
\begin{equation}
l(h,k) = 2^h - 2^{h-k}.
\end{equation}
In particular the longest domain that can be relaxed via a path of
height $h=n+1$ is given by $l(n+1,1)=l(n,n)+1=2^{n}$ from which we
deduce~(\ref{hierarchy}). Related results on the number of
configurations reachable at or below height $h$ can be found in
Ref.~\cite{ChuDiaGra01}.

\subsection{Solution of the dynamics}
\label{sec:solution}

From the result~(\ref{hierarchy}) it follows that the coarsening
dynamics naturally divides into stages distinguished by $n=h(d)-1=0,
1, \ldots$ During stage $n$, the domains with lengths $2^{n-1}<d\leq
2^{n}$ disappear; we call these the `active' domains. This process
takes place on a timescale of
$\order(\rate^{-1}(d))=\order(\eps^{-n})$; because the timescales for
different stages differ by factors of $1/\eps$, we can treat them
separately in the limit $\eps\to 0$. Thus during stage $n$ active
domains are eliminated. The distribution of inactive domains
($d>2^n$) changes because elimination of an active domain implies
coalescence with a neighbouring domain on the right; since the
smallest active domains have length $d=2^{n-1}+1$, any new domain
will have length $\geq 2(2^{n-1}+1)>2^n$ and thus be inactive.

Let $N(d,t)$ be the number of domains of length $d$ at time $t$,
$N(t)=\sum_d N(d,t)$ the total number of domains and
$P(d,t)=N(d,t)/N(t)$ the domain size distribution. Then for a general
process of coarsening by coalescence one has, using that there are no
spatial correlations between domains
\bea
\ddt N(d,t) &=& -\rate(d)N(d,t) - N(d,t)\sum_{d'}\rate(d')P(d',t)
\nonumber\\
& &{}+{} \sum_{d'} N(d-d',t)\rate(d')P(d',t) 
\label{dNdt}
\eea
The first term accounts for disappearance of a domain by coalescence
with its right neighbour, the second for coalescence with a domain on
the left of length $d'$, and the third for creation of larger domains
during coalesence. Summing\eq{dNdt} over $d$ one finds $\partial
N(t)/\partial t
=-\sum_{d}\rate(d)N(d,t)$, and thence
\be
\ddt P(d,t) = -\rate(d)P(d,t) + \sum_{d'} P(d-d',t) \rate(d')P(d',t) 
\label{dPdt}
\ee

We now apply this general result to the scenario at hand. We change to
a rescaled time variable appropriate to the $n$-th stage of the
dynamics, $\tau=t\eps^n$. To be in the $n$-th stage, $t$ has to obey
the restrictions $\eps^{-(n-1)}\ll t \ll \eps^{-(n+1)}$, giving
$\eps\ll\tau\ll\eps^{-1}$; in the limit $\eps\to 0$ we can thus let
$\tau$ range over $[0,\infty]$. The rescaled rates
$\ratee(d)=\eps^{-n}\rate(d)$ are non-zero only for the domains
with $d\leq 2^n$; on the other hand, $P(d,\tau)$ is nonzero only for
$d>2^{n-1}$ since smaller domains have disappeared during earlier
stages. Thus the sum over $d'$ the r.h.s.\ of\eq{dPdt} is restricted to
$2^{n-1}<d'<2^n$, \ie\ the active domains. Furthermore,
if $d$ is an active domain
then $d-d'<2^{n-1}$ and hence the factor $P(d-d',t)$ vanishes for
every term in the sum, reflecting the fact that active domains cannot be
created. Eq.\eq{dPdt} thus becomes for $2^{n-1}<d\leq 2^n$
\be
\ddtau P(d,\tau) = -\ratee(d)P(d,\tau)
\label{dPdt1}
\ee
For $d>2^n$, on the other hand, the {\em first} term in\eq{dPdt} does not
contribute and one gets
\be
\ddtau P(d,\tau) = \sum_{2^{n-1}< d'\leq 2^n} P(d-d',t) \ratee(d')P(d',t)
\label{dPdt2}
\ee
Combining the last two results one finds that the rates drop out,
giving for $d>2^n$
\be
\ddtau P(d,\tau) = \sum_{ 2^{n-1}< d'\leq 2^n }
P(d-d',\tau)\,\left[-\ddtau P(d',\tau)\right]\; .
\label{eqn_motion}
\ee
There is one subtlety in this derivation which we have so far ignored.
Even in terms of the rescaled time $\tau$, \ie\ taking into account
only those relaxation processes which cross the minimum energy barrier
of $n$, the `survival probability' for a domain not to have coalesced
with its right neighbour is not a single exponential which can be
characterized by a rate constant $\ratee(d)$. This is because several
relaxation process with different rescaled rates of $\order(1)$ may
exist. Thus in~(\ref{dPdt1},\ref{dPdt2}) one should replace
$\ratee(d)$ by the negative derivative of the survival probability,
resulting in an effective, time-dependent rate $\ratee(d,\tau-\tau')$
measured from the time $\tau'$ when the domain was created. As
explained above, however, domains that are active cannot be created in
the current stage; thus $\tau'$ must correspond to a previous stage,
which in the limit $\eps\to0$ implies $\tau'\to
0$. Thus~(\ref{dPdt1},\ref{dPdt2}) are correct if $\ratee(d)$ is
replaced by $\ratee(d,\tau)$, and\eq{eqn_motion} follows as before.
Note that, independently of the details of its derivation,
equation\eq{eqn_motion} has a simple intuitive meaning. It expresses
the fact that active domains eliminated during the $n$-th stage
coalesce with neighbouring domains of random size, forming stable
domains of size $d>2^n$. Equation\eq{eqn_motion} then determines the
time dependence of the size distribution of the inactive domains for a
given time dependence of the distribution of the active domains.

To solve the equation of motion\eq{eqn_motion} we define the
generating function
\begin{equation}
G(z,\tau)=\sum_{d=2^{n-1}{+}1}^{\infty} P(d,\tau)z^d
\end{equation}
and its analogue for the active domains,
\begin{equation}
  H(z,\tau)=\sum_{d = 2^{n-1}+1}^{2^n} P(d,\tau)z^d\;.
\end{equation}
Then multiplying (\ref{eqn_motion}) by $z^d$ and summing
over  $d >2^n$ yields
\begin{eqnarray}
  \ddtau \lefteqn{
\left[ G(z,\tau) - H(z,\tau) \right]}\nonumber\\
&=&   \sum_{d=2^{n}{+}1}^{\infty}\, \sum_{d'=2^{n{-}1}{+}1}^{2^{n}}
   P(d-d',\tau) \left[ -\ddtau P(d',\tau)\right] z^d \nonumber\\
%&=&  \sum_{d'=2^{n{-}1}{+}1}^{2^{n}} \sum_{d'=2^{n}{+}1-d'}^{\infty}\, 
%   P(d',\tau) \left[ -\ddtau P(d',\tau)\right] z^d \nonumber\\
&=&  \sum_{d''=2^{n{-}1}{+}1}^{2^{n}} \sum_{d'=2^{n-1}{+}1}^{\infty}\, 
   P(d'',\tau) \left[ -\ddtau P(d',\tau)\right] z^{d''+d'} \nonumber\\
  &=& -G(z,\tau) \ddtau H(z,\tau)
\label{Geqn}
\end{eqnarray}
where we have used that in stage $n$ of the dynamics
$P(d'',\tau)= 0$ for $d''\leq 2^{n{-}1}$. We may
integrate~(\ref{Geqn}) by rewriting it as
\begin{equation}
\frac{1}{\left[1-G(z,\tau)\right]}\ddtau G(z,\tau) =
\ddtau H(d,\tau)
\end{equation}
giving
\begin{equation}
\frac{1-G(z,\tau)}{1-G(z,0)}
=\exp\left[-H(z,\tau)+H(z,0)\right]\;.
\end{equation}
At the end of stage $n$, corresponding to $\tau \to \infty$, all
domains that were active during that stage have disappeared, and so
$H(z,\infty)=0$. Thus
\begin{equation}
G(z,\infty)-1 = [G(z,0)-1]\exp[H(z,0)]\;.
\label{Gresult}
\end{equation}
Now recall that we are considering stage $n$ of the dynamics.  The initial
condition for stage $n+1$ of the dynamics will be given by the
distribution $P(d,t)$ at the end of stage $n$.  Thus defining $G_n(z)
\equiv G(z,0)$ for stage $n$, with a similar definition for the active
generating function $H_n(z)$, Eq.\eq{Gresult} relates the different
stages of the dynamics through
\be
G_{n+1}(z)-1 = [G_n(z)-1]\exp[H_n(z)]\;.
\label{main}
\ee
\begin{figure}
%
%\hspace*{-1mm}
\begin{center}
\epsfig{file=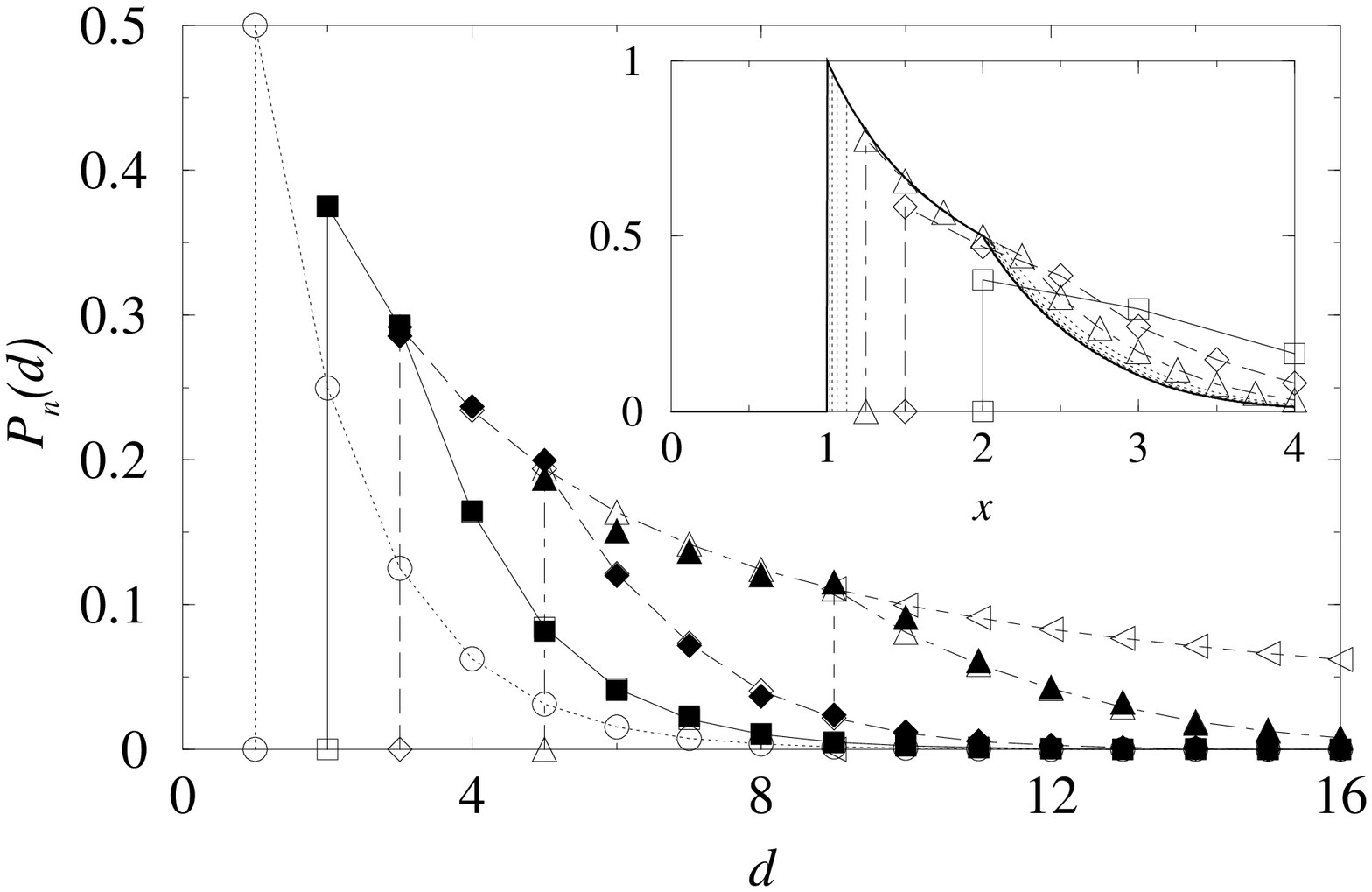, width=8.4cm}
\end{center}
\caption{Domain length distributions $\p_n(d)$ at the end of stage
  $n-1$ of the low $T$ coarsening dynamics, for initial temperature
  $T=\infty$. Open symbols and lines: Theoretical results, calculated
  from~(\protect\ref{main}), for $n=0$ ($\bigcirc$; initial
  condition), 1 ($\Box$), 2 ($\Diamond$), 3 ($\triangle$). Full
  symbols: Simulation results for a chain of length $L=2^{15}$ and
  $\eps=10^{-4}$ ($n=1, 2$) and $\eps=10^{-3}$ ($n=3$). Inset: Scaled
  predictions $2^{n-1}\p_n(d=2^{n-1}x)$ vs.\ $x$ for $n=1, \ldots, 8$.
  Bold line: Predicted scaling function~(\protect\ref{scaling_limit}).
\label{fig:pd}
}
\end{figure}
This exact result connects, through their generating functions,
$\p_n(d)$ and $\p_{n+1}(d)$ which are defined as the domain length
distributions at the end of stages $n-1$ and $n$ of the dynamics
respectively.
It can be checked that for $n=0$, where the system does not cross
energy barriers, Eq.\eq{main} gives results compatible with the known
exact solution of the $T=0$ dynamics~\cite{CriRitRocSel00}.

Iterating\eq{main} from a given initial distribution $\p_0(d)$ gives
$\p_n(d)$ for all $n=1, 2, \ldots$ Fig.~\ref{fig:pd} shows numerical
results for the case where $\p_0(d)$ is the equilibrium
distribution\eq{pd_equil} corresponding to an initial temperature of
$T=\infty$.  It is clear that a scaling limit emerges for large $n$,
\ie\ that the rescaled distributions
\begin{equation}
\ptilde_n(x)=2^{n-1}\p_n(d)
\quad\mbox{where}\quad
x=\frac{d}{2^{n-1}}
\end{equation}
converge to a limiting distribution $\ptilde(x)$ for the scaled domain
size $x$.  This is just a statement of the invariance of the
coarsening processes in each stage once the domain sizes are rescaled
by the characteristic size $2^{n-1}$.

The change to a continuous variable $x$ for the domain lengths simply
results in the generating functions $G_n(z), H_n(z)$ being replaced by
Laplace transforms $g_n(s)$ and $h_n(s)$, via the correspondence
$z^d\to {\rm e}^{-sx}$. Invariance of the scaled domain distribution
under\eq{main} then gives the equation
\begin{equation}
g(2s)-1 = [g(s)-1]\exp[h(s)]\;
\label{mainLap}
\end{equation}
where
\[
  g(s) = \int_1^\infty dx\, \ptilde(x)\, {\rm e}^{-sx}
,\quad h(s) = \int_1^2 dx\,\ptilde(x)\, {\rm e}^{-sx} \;.
\]
We found a solution to\eq{mainLap} by noting that the numerics
strongly suggest $\ptilde(x)=1/x$ for $1<x<2$. Using this as an
ansatz implies
\[
h(s) = \mbox{Ei}(s) - \mbox{Ei}(2s)\quad\mbox{where}
\quad \mbox{Ei}(s) = \int_s^\infty\frac{{\rm e}^{-u}}{u}\,du
\]
which when inserted into (\ref{mainLap}) yields
\[
\left[1- g(s)\right] \exp [\mbox{Ei}(s)] = \mbox{const.}
\]
The requirement that $g(s) \to 0$ for large $s$ fixes the constant as
unity, giving
\be
g(s) = 1-\exp[- \mbox{Ei}(s)]\;.
\label{g_sol}
\ee
Expanding the exponential as a series allows the Laplace transform to
be inverted term by term and one obtains 
\bea 
\ptilde(x)&=&
\frac{1}{2\pi i}\int_{\gamma-i\infty}^{\gamma+i \infty}
ds\,{\rm e}^{sx} g(s) \nonumber \\
&=& \frac{1}{2\pi i}\int_{\gamma-i\infty}^{\gamma+i \infty}
ds\,{\rm e}^{sx} \sum_{m=1}^\infty \frac{(-1)^{m+1}}{m!}{\rm Ei}^m(s)
\nonumber \\
&=& \sum_{m=1}^\infty \frac{(-1)^{m+1}}{m!} \int_1^\infty\!
\prod_{j=1}^{m}\frac{du_j}{u_j} \ 
\delta\left(\sum_{k=1}^{m}u_k-x\right) \nonumber
\\
&=&\Theta(x-1)\,\frac{1}{x}-\Theta(x-2)\,\frac{\ln(x-1)}{x}+\ldots
\label{scaling_limit}
\eea
where $\Theta(x)$ is the Heaviside step function. This result for the
scaling function is shown in Fig.~\ref{fig:pd} and shows good
agreement with the numerically calculated $P_n(d)$ for large $n$.
Note that the series~(\ref{scaling_limit}) has singularities in the
$k$-th derivative at the integer values $x=k+1$, $k+2$, \ldots, a fact
whose physical origin we discuss further in the next subsection.  The
average domain length in the scaling limit is given by
$\dbar_n=2^{n-1}\xbar$; from the results for $\ptilde(x)$ we find
$\xbar=\exp(\gamma)=1.78\ldots$, where $\gamma$ is Euler's constant.

\subsection{A family of related coarsening models}
\label{sec:paste-all}

Surprisingly, the scaling function\eq{scaling_limit} for coarsening
dynamics in the East model is identical to that for the `paste-all'
model of coarsening dynamics, where at each step in the dynamics the
shortest domain is selected and `pasted' onto its left or right
neighbour~\cite{DerGodYek91}.  To understand this, we decide for
definiteness that domains are to be pasted to the right. Then the
paste-all model can be obtained from a modification of the
hierarchical coarsening dynamics discussed so far---one merely needs
to assume that the rates $\rate(d)$ are now {\em all} well separated
from each other, so that at any given stage in the dynamics only
domains of a single size $d$ are active. It is then natural to
consider a family of models which interpolates between paste-all and
the East model, by assuming that the active domain sizes $d$ at any
stage $n$---whose coalescence rates $\rate(d)$ are comparable to each
other but well separated from all other rates---are those with
$a^{n-1}<d\leq a^n$, with $a$ a constant in the range $1<a\leq 2$.

All of the arguments in the previous section apply to this modified
model if powers of $2$ are replaced by powers of $a$ in the
appropriate places. In particular, Eq.\eq{mainLap} for the Laplace
transform of the scaling distribution becomes
\begin{equation}
g(as)-1 = [g(s)-1]\exp[h_a(s)]
\label{modLap}
\end{equation}
with
\be
h_a(s) = \int_1^a dx\,\ptilde(x)\, {\rm e}^{-sx}
\label{ha}
\ee
generalizing the earlier $h(s)\equiv h_2(s)$. We now show
that\eq{modLap} has the same family of solutions for any $a\leq 2$,
thus proving in particular the link between the scaling distributions
of the East model ($a=2$) and the paste-all model ($a\to 1$). To see
this, consider first $a\to 1$; we should then recover the scaling
equation for the paste-all model. Indeed, setting
$a=1+\delta$ and expanding to first order in $\delta$, Eq.\eq{modLap}
becomes
\begin{equation}
sg'(s) = [g(s)-1]\ptilde(1){\rm e}^{-s}
\label{paste_all}
\end{equation}
which is equivalent to Eq.~(38) in Ref.~\cite{DerGodYek91}.
Integrating\eq{paste_all} gives~\cite{DerGodYek91}
$g(s)=1-\exp[-\ptilde(1){\rm Ei}(s)]$, and the Laplace transform can
be inverted as in\eq{scaling_limit}, giving
\[
\ptilde(x)=
\Theta(x-1)\,\frac{\ptilde(1)}{x}-\Theta(x-2)\,\frac{\ptilde^2(1)\ln(x-1)}{x}
+\ldots
\]
Given that only the first term contributes for $x\leq 2$, it is now
easy to verify that this is also a solution of\eq{modLap} 
for any $a\leq 2$: from\eq{ha} one has $h_a(s)=-\ptilde(1)[{\rm
Ei}(as)-{\rm Ei}(s)]=\ln[1-g(as)]-\ln[1-g(s)]$.

In summary, we have shown that there is a whole family of coarsening
models which interpolate between the paste-all and East models and
which have the same scaling solution. The solution is, in principle,
parameterized by $\ptilde(1)$ but, as discussed in~\cite{DerGodYek91},
the requirement of a finite mean for the scaling distribution imposes
$\ptilde(1)=1$. Thus $g(s)=1-\exp[-{\rm Ei}(s)]$ as we found
in\eq{g_sol}.

Physically, the common feature of all the models in the family is that
active domains can never be created during the stage where they are
active; this breaks down for $a>2$ where indeed the scaling solution
no longer applies. This argument clarifies the origin of the
singularities at integer arguments of $\ptilde(x)$: the existence of a
shortest scaled domain length $x=1$ implies that the shortest inactive
domain that can be created has $x=2$, and this effect then propagates
to $x=3,4,\ldots$ as the dynamics is iterated. We initially thought
that the fact that in the East model the active domains cover a range
$x\in[1,2]$ bounded by two integers played a role, but this is
clearly not so since for general $a \leq 2$ this range becomes
$x\in[1,a]$.

\subsection{Anomalous coarsening}

Having solved the dynamics in terms of the different stages labelled
by $n$, we now translate these results into actual time dependencies.
As before, we consider a system quenched from, say, $T =\infty$, to a
temperature $T \ll 1$ at time $t=0$. If data for \eg\ the average
domain length $\bar{d}$ are plotted against the scaled time variable
$\logtime= T\ln t$, then for $T\to 0$ the $n$-th stage of the dynamics
shrinks to the point $\logtime=n$. In this limit we predict that, for
$n-1<\logtime<n$, the domain length distribution is $\p_n(d)$ as
defined by the recursion\eq{main}. The average domain length
$\bar{d}(\logtime)$ will follow a `staircase' function, jumping at
$\logtime=n$ from $\bar{d}_n=\sum_d \p_n(d)d$ to $\bar{d}_{n+1}$. This
was illustrated and verified by low-$T$ simulations
in~\cite{SolEva99}.

For large $n$, \ie\ in the
scaling regime of large $\bar{d}$ (but still $\bar{d}\ll \deq$) we
know that $\dbar_n=2^{n-1}\xbar$ where $\xbar=1.78\ldots$ The
staircase function $\bar{d}(\logtime)$ is therefore bounded between
$2^{\logtime-1}\xbar \leq \dbar\leq 2^{\logtime}\xbar$, giving
$\half\leq \dbar/(\xbar\, t^{T \ln 2}) \leq 1$ when expressed in terms
of ordinary time $t$. This shows that the hierarchical dynamics gives
rise to anomalous coarsening, \ie\ the typical domain size grows as
$\dbar \sim t^{T \ln 2}$, more slowly in time than a usual power law
with $T$-independent exponent. The exponent can also be deduced from
the scaling $t\sim \rate(\dbar)^{-1} \sim \eps^{n} = \exp(n/T)$ of the
time required to eliminate a domain of the typical size $\dbar$. Since
$\bar{d}\sim 2^n$ asymptotically, one has $n\approx \ln \bar{d}/\ln 2$
and this time scales as $t\sim \bar{d}^{1/T\ln 2}$, giving $\dbar \sim
t^{T \ln 2}$ as before. For decreasing $T$ the coarsening becomes
anomalously slow and in fact logarithmic for $T\to 0$.

By extrapolating the anomalous coarsening law to the equilibrium
domain length $\deq=\exp(1/T)+\order(1)$, we can estimate the
equilibration time of the system for $T\to 0$ as
$\teq\sim\tst=\exp(1/T^2 \ln 2)$. Of course, in performing this
extrapolation, we are using the anomalous coarsening law in the regime
where domain sizes are $\order(1/\eps)$ and the assumption of
irreversible down-flips no longer holds. Given this we could have
equally well extrapolated $\dbar \sim t^{T\ln 2}$ to some multiple
$\alpha\deq$ of the equilibrium domain length, which would yield $\teq
=\exp(1/T^2\ln 2-A/T)$ with $A=-\ln\alpha/\ln 2$. With the sign as
defined, $A$ should be positive, corresponding to $\alpha<1$; a
negative value of $A$ is excluded as it would violate the relaxation
time bound of~\cite{AldDia02}. Thus we expect $\tst$ to give only the
dominant divergence of the equilibration time $\teq$, with subdominant
correction factors $\exp(-A/T)$ making $\teq\ll\tst$ generically. Our
analysis in Sec.~\ref{sec:eq} suggests, on the other hand, that the
longest relaxation timescales {\em at} equilibrium are given directly
by $\tst$. This is not unreasonable: the time to reach equilibrium
from an initial high temperature configuration is expected to be much
shorter than the time for correlations to be erased starting from an
equilibrium (low temperature) initial condition, simply because the
typical domains to be eliminated are shorter. 

Finally, we comment
briefly on the implications of the coarsening dynamics for the
two-time spin autocorrelation function $C(t,\tw)=\lav s_i(t)
s_i(\tw)\rav - \lav s_i(t)\rav \lav s_i(\tw)\rav$
where $t>\tw$.  Since in the limit
$T\to 0$ spins flip down irreversibly, 
$s_i(t)=1$ implies that $s_i(\tw)=1$ at the earlier
time $\tw$, thus
$C(t,\tw)=c(t)[1-c(\tw)]$ where $c(t)$ is the time-dependent up-spin
concentration. As a function of $t$, $C(t,\tw)$ will thus exhibit the
same plateaux as $c(t)$, and these were indeed observed in the
simulations of~\cite{CriRitRocSel00}. Also, the normalised
autocorrelation becomes simply
$C(t,\tw)/C(\tw,\tw)=c(t)/c(\tw)=\dbar(\tw)/\dbar(t)$. This dependence
on only the ratio of the relevant lengthscales at times $\tw$ and $t$
is natural for a coarsening process, and gives a good fit to the data
of~\cite{CriRitRocSel00}.

\section{Equilibrium dynamics}
\label{sec:eq}

\subsection{Functions of interest}
\label{sec:functions}

We now turn to the equilibrium dynamics of the East model at low
temperatures. These are encoded in relaxation functions such as
correlations and spin persistence (see below);
response functions provide no new information since they are related
to correlations via the fluctuation-dissipation theorem (FDT). For
correlation functions standard choices would be the local spin
autocorrelation $\langle s_i(t) s_i(0)\rangle-c^2$ or the correlation
function of the up-spin concentration (or magnetization)
$M(t)=(1/L)\sum_i s_i(t)$. However, the directionality of the kinetic
constraint can be shown to imply that all non-local correlations
$\langle s_i(t) s_j(0)\rangle-c^2$ vanish~\cite{JaecSap93} so these
two choices give the same information. We can thus focus on the
normalized autocorrelation function
\be
C(t)=\frac{\langle s_i(t)s_i(0)\rangle-c^2}{c-c^2} =
\frac{P(s_i(t)=1|s_i(0)=1)-c}{1-c}
\label{Cdef}
\ee
This decays from $C(0)=1$ to $C(t\to\infty)=0$; the second equality
in\eq{Cdef} implies that it is essentially the conditional probability
for an up-spin that was up at time 0 also to be up at time
$t$. Closely related is the up-spin persistence function $\pup(t)$,
which gives the probability that a spin is up throughout the time
interval $[0,t]$, and the analogously defined down-spin
persistence $\pdown(t)$.

Fortunately, it turns out that $C$, $\pup$ and $\pdown$ do not need to
be analysed separately because they all become identical in the limit
$T\to 0$ (taken at a constant value of, say, $C(t)$) that we are
interested in. To see this, note that since in the East model the
spins are only coupled via the kinetic constraint, which acts to the
right, the dynamics of a spin $s_i$ cannot influence that of its left
neighbour $s_{i-1}$. (This is strictly true only in the limit of an
infinite chain $L\to\infty$, but that is precisely the case of
interest.) Furthermore, the dynamical evolution of spin $s_i$ is that
of a single spin in a field whenever $s_{i-1}=1$, and is completely
frozen otherwise. It follows that if, between times $0$ and $t$,
$s_{i-1}$ has been up a fraction $m$ of the time, then the evolution
of $s_i$ has been that of a single spin over time $mt$. But the
single-spin correlation and persistence functions are trivial to work
out; denoting the distribution of $m$ for a given $t$ by $P(m;t)$, one
thus has the simple expressions
\bea
C(t)      &=& \int_0^1 dm\,P(m;t)\,{\rm e}^{-(1+\eps)mt} \label{Cm} \\
\pup(t)   &=& \int_0^1 dm\,P(m;t)\,{\rm e}^{-mt} \label{P1m} \\
\pdown(t) &=& \int_0^1 dm\,P(m;t)\,{\rm e}^{-\eps mt} \label{P0m}
\eea
It follows trivially that 
\be
C(t)\leq\pup(t)\leq\pdown(t)
\label{basic_ineq}
\ee
Now since $P(s_i(t)=1|s_i(0)=1) \geq \pup(t)$ (the probability for a
spin to be up at times $0$ and $t$ must be greater than that for it to
be up at $0$ and $t$ {\em and} all intermediate times), one has
from\eq{Cdef}
\be
(1-c)C(t)+c\geq \pup(t)
\label{C_P1_ineq}
\ee
Eqs.~(\ref{basic_ineq},\ref{C_P1_ineq}) together show that
$C(t)$ and $\pup(t)$ coincide in the low-$T$ limit ($c\approx \eps\to
0$) whenever the values of the functions themselves are of
$\order(1)$.

To get the analogous result for $C$ and $\pdown$, consider
$P(m;t)$. The concentration of persistent up-spins and down-spins in
the system is $c\pup(t)$ and $(1-c)\pdown(t)$, respectively, and these
have $m=1$ and $m=0$. Thus
\be
P(m;t)=(1-c)\pdown(t)\delta(m)+c\pup(t)\delta(m-1)+\ldots
\label{Pm_t}
\ee
where the dots indicate contributions for $m$ strictly between 0 and
1. Inserting into\eq{P1m} and separating off the $\delta(m)$-term
gives
\[
\pup(t) = (1-c)\pdown(t) + \int_{>0}^1 dm\,P(m;t){\rm e}^{-mt} \geq
(1-c)\pdown(t)
\]
Together with\eq{basic_ineq} it follows that also $\pup(t)$ and
$\pdown(t)$ coincide for $c\to 0$. Thus $C=\pup=\pdown$ in the limit,
and we can restrict attention to \eg\ the up-spin persistence
$\pup(t)$ in the following.

We stress once more that the directionality of the kinetic constraint
is essential for the above arguments to work. In the undirected
version of the East model, \ie\ the one-dimensional
Fredrickson-Andersen model with $f=1$, it would still be true that the
dynamics of spin $s_i$ is determined by the amount of time which
$s_{i-1}$ (and $s_{i+1}$) spend in the up-state. But this time cannot
be determined independently of the state of spin $s_i$, since $s_i$
itself affects the dynamics of $s_{i-1}$ (and $s_{i+1}$).

\subsection{First passage times}
\label{sec:FPT}

%Things to cover:
%
%explanation of hypothesis for relaxation time - distance scaling,
%introduce scaled time $\dt$ and distance $\dd$, argue that up- and
%down-flip FPTs should be asymptotically equal, and that fluctuations
%in rescaled time should vanish in the limit. Show some numerical FPT
%data for support. Conclude that have continuous timescale separation
%(order of limit: always considering fixed $\dd_1$ and $\dd_2$ and then
%$T\to 0$). Use Diaconis result to conclude that for $\dd\to\infty$
%should get $\dt\to 1$ (or at least $\leq 1$). Explain possibilities
%for shape of limiting plot and why a completely flat bit would be
%implausible (or maybe do this later once have superdomain model
%results).

To get some insight into the equilibrium dynamics, we generalise the
domain coalescence rates $\rate(d)$, discussed in 
Sec.~\ref{sec:noneq}  for $d=\order(1)$
and $\epsilon\to 0$, to the regime of domain sizes typically found in
equilibrium, $d\sim\deq \approx 1/\epsilon$.  Starting from the state
$(s_0,s_1, \ldots, s_d)$ = $10\ldots 01$, we define a mean first
passage time (MFPT) $\mfpt(d)$ as the mean time for the spin $s_d$ to
flip down for the first time, and we set
$\rate(d)=\mfpt^{-1}(d)$. Note that in Sec.~\ref{sec:barriers} we
required for a `first passage' that not just $s_d$ but all other spins
$s_1,\ldots, s_{d-1}$ be down; here we just require that $s_d=0$.  In
the regime considered in Sec.~\ref{sec:noneq}, the two definitions are
equivalent to leading order, since the relaxation process has passed
its `highest point' once $s_d$ has flipped down, resulting in the same
energy barrier whether or not the process continues to the empty
state.
% The scaling of $\rate(d)\sim \eps^n$ thus remains
%valid for $2^{n-1}<d\leq 2^n\ll 1/\eps$.

\begin{figure}
\epsfig{file=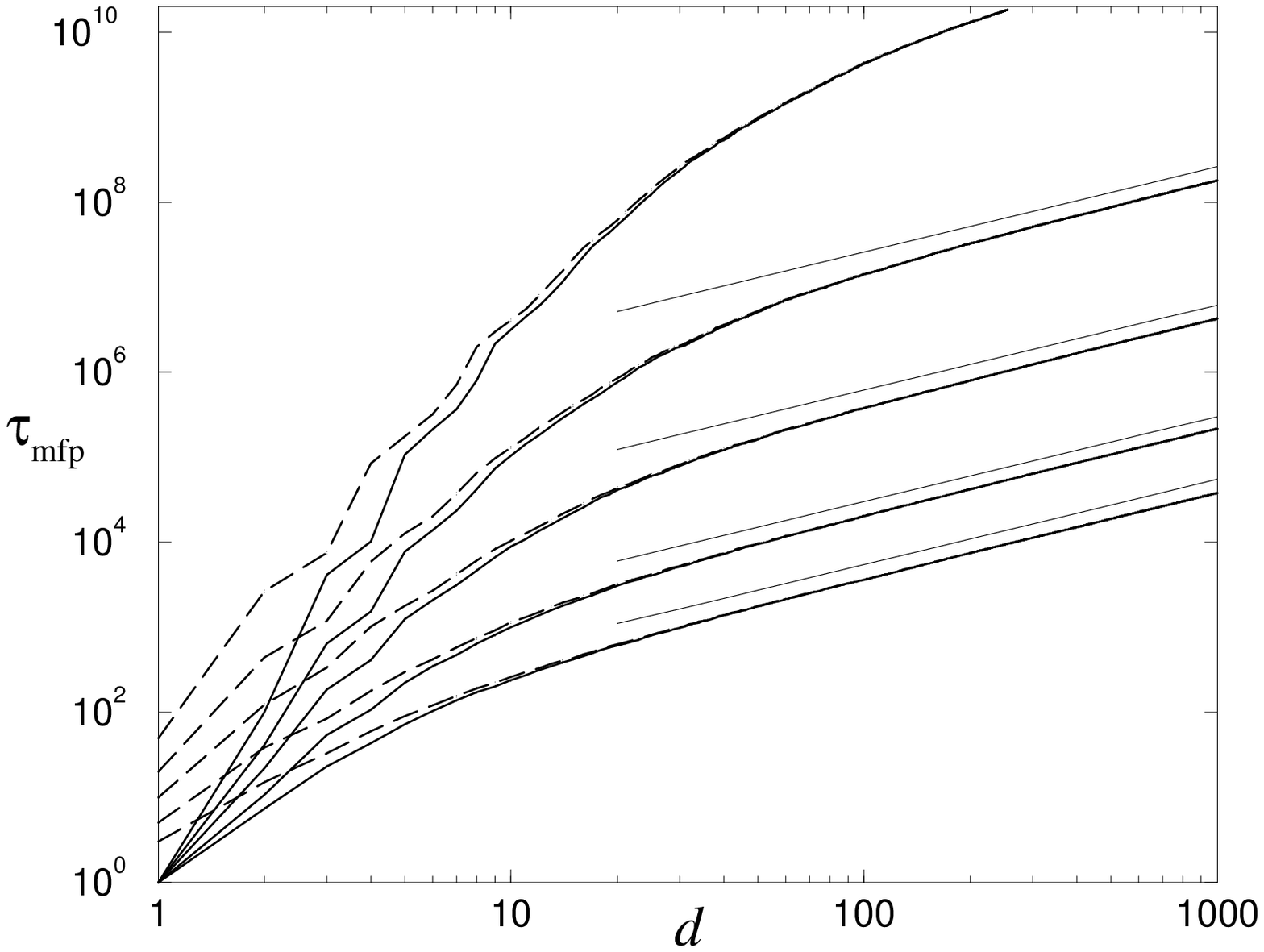,width=8.4cm}
\caption{Simulated MFTPs $\mfpt$ for down-flips (solid) and $\mfpt\up$
for up-flips (dashed) as a function of unscaled distance $d$, for
$\eps=0.3, 0.2, 0.1, 0.05, 0.02$ (bottom to top). The thin solid lines
have slope 1, showing that for large $d$ the MFPTs vary linearly with
$d$. Error bars are too small to show on the scale of the plot ($\leq
3.5\%$ relative error on $\mfpt$ and $\mfpt\up$).
\label{fig:raw_MFPTs}
}
\end{figure}

Consider now the dependence of $\mfpt(d)$ on $d$. From the discussion
of the out-of-equilibrium dynamics we know that $\mfpt(d)$ exhibits a
step-like variation with $d$ for $d=\order(1)$ and small $\eps$,
varying only by factors of order unity within each range
$2^{n-1}<d\leq 2^n$ but increasing by $\sim\eps^{-1}$ between
ranges. As $n$ increases at fixed $\eps$, however, the $\order(1)$
changes within ranges become larger and eventually comparable to
$\eps^{-1}$; $\mfpt(d)$ will then increase more smoothly with $d$.
The numerically simulated MFPTs in Fig.~\ref{fig:raw_MFPTs} confirm
this. Finally, for $d$ larger than $\deq\approx 1/\eps$, the figure
shows that the MFPTs increase linearly with $d$. Intuitively, one then
has a `front' of up-spins that creeps `ballistically' along the
chain. This can be motivated by arguing that, for $d$ above the
typical equilibrium lengthscale $\deq$, the spin front no longer
remembers that it originated from a single up-spin a distance $d$ to
the left. Instead it behaves as on an infinite chain, where its
propagation velocity is necessarily constant.

\begin{figure}
\epsfig{file=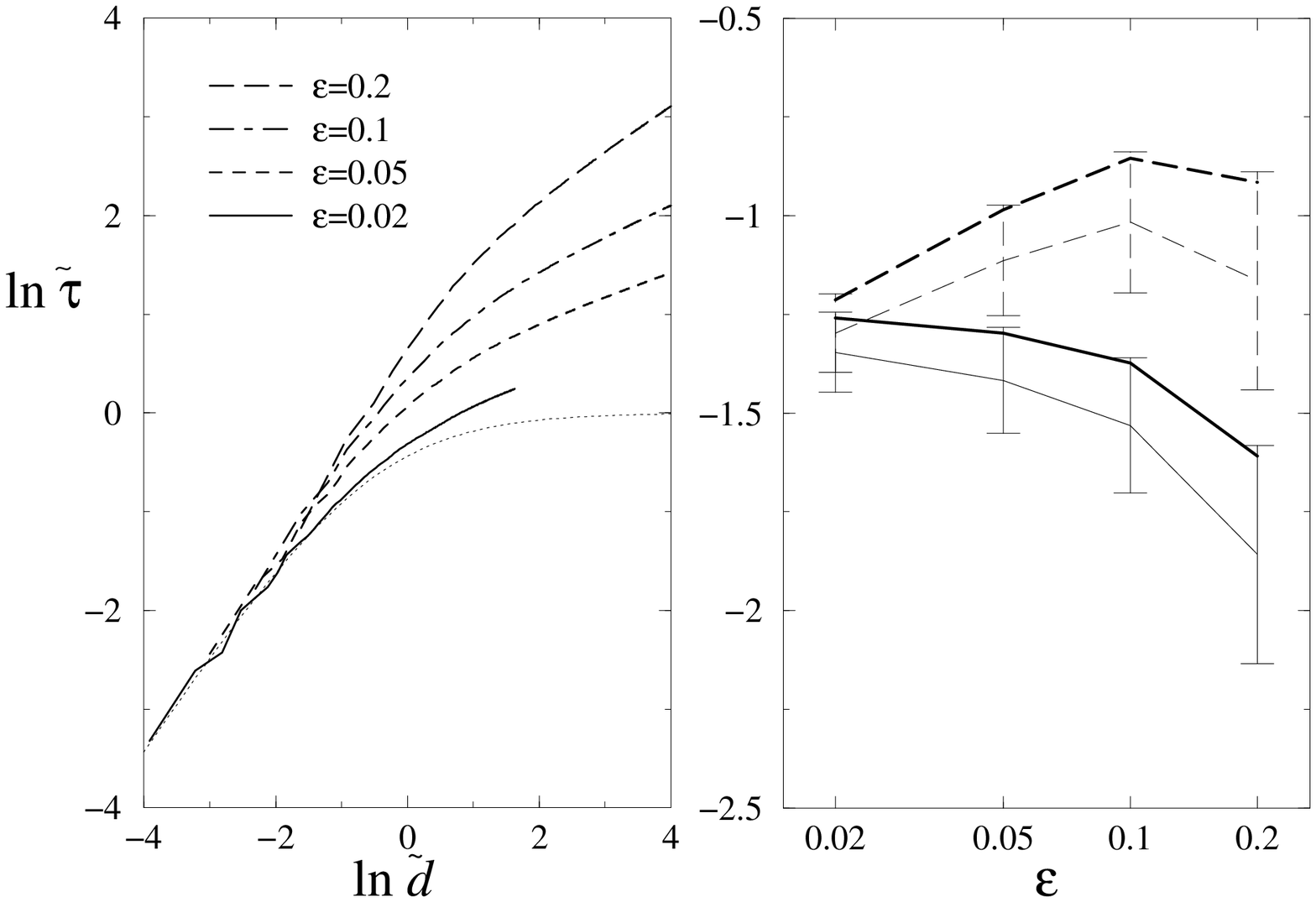,width=8.4cm}
\caption{Left: Logarithmic plot of scaled FPTs $\dtau$, as defined
in~(\protect\ref{timescaling}), versus scaled distance $\dd$, for
$\eps=0.2, 0.1, 0.05, 0.02$ (top to bottom on right). The dotted line
shows a qualitative sketch of the scaling function $\scf(\dd)$, with
$\scf\to 1$ ($\ln\scf\to 0$) for $\dd\to\infty$; see text. Right:
Up-flip (dashed) versus down-flip (solid) FPTs and their fluctuations,
versus $\eps$ for constant $\dd=\eps d=0.1$. Thin lines and error
bars: mean and $\pm\half$(standard deviation) of the log-rescaled FPT
$\ln\dtau$. Thick lines: $\ln \mfptsc$. As $\eps$ decreases, the
fluctuations in $\ln\dtau$ are seen to decrease: error bars shrink,
and $\ln\mfptsc$ and the mean of $\ln\dtau$ become closer. Also,
up-flip and down-flip times become closer; both observations support
our scaling hypothesis for the FPTs.
\label{fig:scaled_MFPTs}
}
\end{figure}

So far we have considered the variation of $\mfpt(d)$ with $d$ at
fixed $\eps$. One suspects, however, that \eg\ the decay of the
persistence function $\pup(t)$ is governed by a lengthscale $d(t)$,
giving the size of the largest domains that have had time to
equilibrate, and more precisely---since we are considering equilibrium
dynamics---by the ratio $d(t)/\deq$. We thus define scaled domain
sizes $\dd=\eps d$ ($=d/\deq$ for $\eps\to 0$) and now ask how
$\mfpt(\dd)$ scales with $T$ for {\em fixed} $\dd$. For small $\dd$,
extrapolating from the regime of $d=\order(1)$, and using $n\approx
\ln d/\ln 2$, one expects
\beastar
\mfpt(\dd) &\sim &\eps^{-\ln d/\ln 2} = \eps^{-\ln\dd/\ln 2 - 1/(T \ln
2)} \\
&=& \exp\left(\frac{\ln\dd}{T\ln 2}+\frac{1}{T^2\ln 2}\right)
\eeastar
Rearranging, one has
\[
\left(\frac{\mfpt(\dd)}{\tst}\right)^{T \ln 2} \sim \dd
%\label{timescaling_aux}
\]
From the downward curvature of the
lines in Fig.~\ref{fig:raw_MFPTs}, one sees that this expression will
not be valid for larger $\dd$, where the r.h.s.\ will cross over to a
slower increase with $\dd$. We thus make a {\em time scaling
hypothesis} which replaces $\dd$ by a more general scaling function
$\scf(\dd)$ of $\dd$, \ie\ we assume that the rescaled MFPT approaches
\be
\mfptsc(\dd)\equiv\left(\frac{\mfpt(\dd)}{\tst}\right)^{T \ln 2} \to
\scf(\dd) \quad\mbox{for}\quad T\to0
\label{timescaling}
\ee
when the limit is taken at fixed
$\dd$. Fig.~\ref{fig:scaled_MFPTs}(left) represents numerical data for
the MFPTs in this scaled form and shows that the
assumption\eq{timescaling} is certainly plausible for not-too-large
$\dd$. For larger $\dd$, convergence to a scaling limit is not yet
evident from the data that we can generate on practical simulation
timescales. This is because for $d \gg \deq$ the ballistic propagation
discussed above implies $\mfptsc\propto d^{T\ln 2} \propto \dd^{T\ln
2}$ which will tend to a constant as $T\to 0$ but does so very
slowly. For the scaling function $\scf(\dd)$ this implies that it must
approach a constant as $\dd\to\infty$; recall that this limit is taken
{\em after} the limit $T\to 0$ at fixed $\dd$. We note that the bounds
of~\cite{AldDia02} imply that the MFPTs $\mfpt$ cannot exceed $\tst$
by more than factors of $\order(1)$; this implies from\eq{timescaling}
that $\scf(\dd)\leq 1$. Our numerical data in
Fig.~\ref{fig:scaled_MFPTs}(left) are restricted to values of $T$
which are still too large to determine $\scf(\dd)$ with any accuracy,
but the tentative extrapolation to $T\to 0$ indicated in the figure is
certainly consistent with the bound $\scf(\dd)\leq 1$.

We work below with a slightly strengthened version
of\eq{timescaling}. The first passage-time $\fpt$ for down-flips is a
fluctuating quantity, with mean $\mfpt$. We assume that fluctuations
in $\fpt$ are small enough so that\eq{timescaling} holds even for the
fluctuating $\fpt$, which means that for $T\to 0$ the rescaled FPT
$\fptsc(\dd)=({\fpt(\dd)}/{\tst})^{T \ln 2}$ becomes
non-fluctuating. This assumption is not as strong as it may sound;
because of the exponentiation by $T\ln 2$ it holds \eg\ if the
relative fluctuations of the unscaled FPT $\fpt(\dd)$ remain of
$\order(1)$ as $T\to 0$. Further confirmation comes from
Fig.~\ref{fig:scaled_MFPTs}(right) which shows numerical data for
$\dd=0.1$.

In a final generalization we also assume that\eq{timescaling} applies
if we consider FPTs $\fpt\up$ for {\em up-flips}, where we start with
the empty state $10\ldots 00$ and $\fpt\up$ is the first time where
the last spin, $s_d$, has flipped up.  Naively one might have
suspected that $\fpt\up$ and $\fpt$ differ by a factor $\sim 1/\eps$.
However, the equivalence of $\pup(t)$ and $\pdown(t)$ that we proved
for low $T$ already suggests otherwise. The correct intuition is that
both up- and down-spins only become mobile once an up-spin `front'
from an up-spin to the left has reached them, and that for $T\to0$ the
time required for this front propagation vastly dominates the effect
of the different flip rates of up- and down-spins once the front has
reached them. Fig.~\ref{fig:scaled_MFPTs}(right) again supports this
assumption with numerical data for $\dd=0.1$.

In summary, we assume in the following that\eq{timescaling} is valid
for unaveraged up-flip and down-flip FPTs, with the same scaling
function $\scf(\dd)$. A consequence of this is {\em continuous
timescale separation}: for any two rescaled distances $\dd_1<\dd_2$,
the ratio of the corresponding FPTs is
$[\scf(\dd_2)/\scf(\dd_1)]^{1/(T \ln 2)}$ and diverges as $T\to 0$. In
the limit, this means that the equilibration of domains of length
$\dd_1$ proceeds infinitely more quickly than for any even slightly
larger length $\dd_2$. This insight is key for the construction of the
superdomain model described next. A proviso is that we have assumed
here that the scaling function $\scf(\dd)$ is strictly monotonically
increasing with $\dd$. In principle, it is possible that $\scf(\dd)$
could increase monotonically only up to some finite $\dd_*$, and be
exactly constant thereafter. This seems to us implausible,
however---\eg\ it is difficult to conceive of a physical mechanism
causing the singularity at $\dd_*$---and, as discussed below, would
also give very unusual predictions for the time-dependence of the
low-$T$ relaxation functions.

\subsection{Superdomain model}

We now exploit the idea of continuous timescale separation to
construct an effective description for the low-$T$ equilibrium
dynamics of the East model. It is natural to work with the
equilibrated lengthscale $\ddr$ that corresponds to the
relaxation timescale $\tr$ we are considering; from\eq{timescaling},
the two are related by
\[
\left(\frac{\tr}{\tst}\right)^{T \ln 2} = \scf(\ddr)
\]
Here and in the following, the limit $T\to 0$ (or equivalently
$\eps\to0$, or $c\to0$) is always understood. For simplicity we will
simply call $\ddr$ `time' where there is no ambiguity. We
stress that we work with {\em rescaled} lengthscales $\dd=d\eps$
throughout; a value of $\dd$ of over unity thus corresponds to a very
large domain length $d=\dd/\eps$ in the limit $\eps\to 0$.

Consider now an equilibrium configuration of the spin chain,
consisting of up-spins separated by long domains of
down-spins. From\eq{pd_equil}, the (rescaled) domain sizes are
distributed according to a simple exponential, $P(\dd)=\exp(-\dd)$. A
`time' $\ddr$ later, each of the up-spins will have developed an
equilibrated zone of length $\ddr$ to its right: from continuous
timescale separation, the FPT to reach any spin {\em within} this zone
is much shorter than $\tr$. Throughout the zone we should thus indeed
have equilibrated spins, which are independently up with probability
$c=\eps/(1+\eps)\approx \eps$. The domains defined by up-spins within
these equilibrated zones will not be of interest in the following.
Instead we focus on the `superdomains' which are naturally defined by
the unequilibrated up-spins, to which we refer as `superspins'.  Thus
each superdomain is bounded on the left by such a superspin, followed
by an equilibrated zone of length $\ddr$ and then a string of
down-spins.

\begin{figure}
\epsfig{file=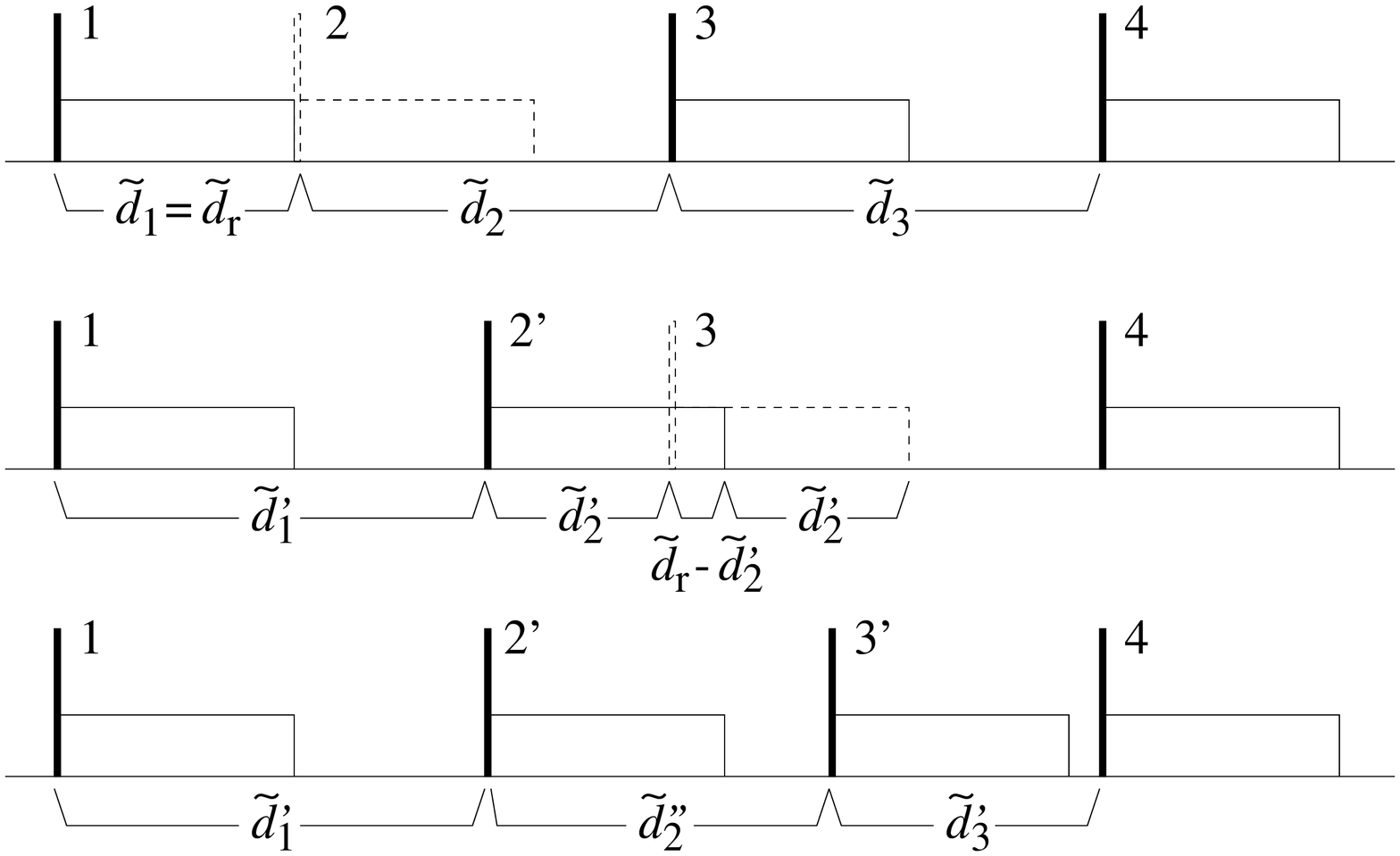,width=8.4cm}
\vspace*{2mm}
\caption{Illustration of superdomain dynamics. Superspins are
shown by the thin vertical rectangles; each has an equilibrated zone
of length $\ddr$ to its right, indicated by a horizontal
rectangle. Shown is a situation where the relaxation `time' $\ddr$ has
just become equal to the smallest superdomain length, $\dd_1$. The
equilibrated zone of superspin 1 then `catches up' with 2 and
equilibrates this spin (first line). At this moment, there can be
up-spins within the former equilibrated zone of $2$; if there are, the
leftmost of these becomes a new superspin $2'$, bounding a superdomain
of length $\dd'_2$ (second line). If $\dd'_2<\ddr$, the newly created
equilibrated zone of $2'$ catches superspin 3. Within the zone of
length $\dd'_2$ that is now no longer equilibrated (dashed) up-spins
can again remain, with the leftmost becoming a new superspin $3'$. In
the example, the relaxation process at $\ddr$ stops at this point
since $\dd'_3>\ddr$; otherwise, it would continue in the same fashion.
\label{fig:superdomain}
}
\end{figure}

To understand the dynamics of superdomains as we look at increasing
times $\ddr$, it is easiest to consider a system for which the
rescaled chain length $\eps L$ and therefore the typical number of
superdomains is large but finite; there is then at all
times a nonzero minimum superdomain length. On increasing $\ddr$ from
zero, all superspins remain as they are as long as the equilibrated
zone of each superdomain has not yet reached the superspin to the
right. However, when $\ddr$ becomes equal to the smallest superdomain
length in the system, $\dd_1$ say, then this superdomain's
equilibrated zone `catches up' with superspin 2 bounding the next
superdomain (of length $\dd_2$) on the right. This superspin now
becomes equilibrated, and so one might assume that the two
superdomains just coalesce, forming a single superdomain of length
$\dd_1+\dd_2$. This, however, will only be the case if there are no
up-spins in the equilibrated zone of superspin 2 at this time; since
this zone has length $\ddr$, the probability for this event is
$\exp(-\ddr)$. Otherwise, \ie\ with probability $1-\exp(-\ddr)$, there
will be at least one up-spin in the equilibrated zone of superspin
2. The leftmost of these, having lost the superspin from which it
became equilibrated, becomes frozen and thus itself turns into a
superspin, $2'$. The distance $\del$ between $2'$ and the old
superspin 2 has probability distribution $\exp(-\del)/[1-\exp(-\ddr)]$
over the interval $\del\in[0,\ddr]$, and the two new superdomains
have length $\dd_1'=\dd_1+\del$ and $\dd_2'=\dd_2-\del$ (see
Fig.~\ref{fig:superdomain}).

What makes the superdomain model nontrivial is that in the second case
the process of superspin elimination and regeneration can now
continue. Superspin $2'$ will immediately (on the timescale $\ddr$
being considered) equilibrate a zone of length $\ddr$. If
$\dd_2'<\ddr$, then this zone includes superspin 3, which now becomes
equilibrated along with a segment of length $\ddr-\dd_2'$ to its
right. This, however, still leaves a segment of length
$\ddr-(\ddr-\dd_2')=\dd_2'$ of the old equilibrated zone of superspin
3, which is now frozen (see Fig.~\ref{fig:superdomain}). If there is
no up-spin within this segment when superspin 3 equilibrates, then
superdomains $2'$ and 3 coalesce; this happens with probability
$\exp(-\dd_2')$. Otherwise, the leftmost up-spin in the segment
freezes into a new superspin $3'$, and the process continues by
iteration. It is clear that only superdomains of length $\dd>\ddr$
have been generated when the process terminates. One can thus now
increase $\ddr$ until the new minimum superdomain size is reached, at
which point a new relaxation process as described above starts.

\begin{figure*}
\begin{center}
\epsfig{file=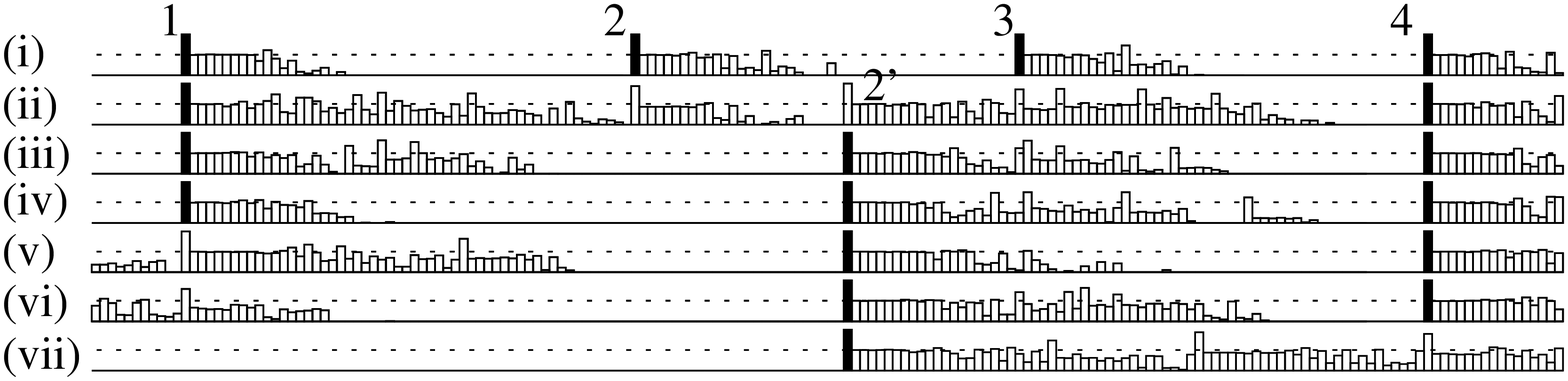,width=17cm}
\end{center}
\vspace*{2mm}
\caption{A direct simulation
run on a system at $\eps=0.02$ which provides
support for the notion of superdomain-type dynamics. Shown
is a section of 180 spins out of a much longer chain, for a single
simulation run. The lines correspond to seven evenly (logarithmically)
spaced time intervals within the range $6.26\times 10^8<t<10^{10}$.
The boxes indicate the current magnetizations $\bar{s}_i$ of the
spins, determined by averaging over the relevant time interval. A
logarithmic scale is used, so that the highest boxes correspond to
$\bar{s}_i=1$, the dashed lines to the equilibrium value
$\bar{s}_i=c$, and the baselines to $\bar{s}_i=c^2$. (Lower
values of $\bar{s}_i$ are not shown.)  Filled boxes indicate spins which are
persistent over the time interval ($\bar{s}_i=1$). Note the similarity
between the first two lines here and the superdomain dynamics sketched
in Fig.~\protect\ref{fig:superdomain}: within the time interval (ii),
the equilibrated zone of superspin 1 catches up with 2; a new
superspin ($2'$) is created and eliminates superspin 3. Lines (v) and
(vii) show events where superspins 1 and 4 are eliminated. Of course,
$\eps$ here is still too large to be in the asymptotic limit $T\to 0$
where superdomain dynamics applies exactly. This is why, in contrast
to Fig.~\protect\ref{fig:superdomain}, the equilibrated zones of the
superspins are not all of the same length and do not have sharply
defined boundaries.
%
% Use lines 5-9 from picture history.ps, including superspin on the
% left. Make all ``high'' boxes black, and all very low ones white.
% Probably ought to label to the superspins. In xmgr, overlaying
% rectangles to get the right section? Stuff is in
% /linear_response/direct_sims/autocorr/programs_scripts/ 
% 7 intervals, logarithmically spaced, between 6.26\times 10^8 and 10^10
% 180 spins
\label{fig:history}
}
\end{figure*}
Above, we arrived at the superdomain model starting from
the hypothesis of continuous timescale separation. The model lets us
access, via an effective description, very low temperatures
corresponding to extremely long relaxation times. By definition it is
therefore difficult to demonstrate superdomain-type dynamics on the
timescales of a numerical simulation. Nevertheless, the sample run at
$\eps=0.02$ shown in Fig.~\ref{fig:history} illustrates some important
features of superdomain dynamics, in particular the regeneration of
superspins and the resulting propagation of equilibrated zones.

To summarize the superdomain model, let us restate its dynamics. 
We present this in the form of a schematic simulation
algorithm; a formal definition of the stochastic evolution of the
sequence of superdomain lengths is of course possible but would be
more awkward. We re-emphasize that all lengths are {\em rescaled}
lengths, $\dd=d\eps$.

\begin{enumerate}

\item Initialize superdomain lengths from an exponential
distribution $P(\dd)=\exp(-\dd)$.

\item Set $\ddr=$ size of smallest current superdomain. Let that size
be $\dd_1$, with $\dd_2,\dd_3,\ldots$ the sizes of the superdomains on
the right. Set $i=1$.

\item Delete the now equilibrated superspin $i+1$. With probability
$\exp(-\dd_i)$, 
%
%there is no up-spin in the part of the equilibrated
%zone of superspin $i+1$ that now becomes frozen. This means that 
%
the superdomains $i$ and $i+1$ coalesce and the relaxation process at
this $\ddr$ is 
complete: let $\dd_i\leftarrow \dd_i+\dd_{i+1}$, delete superdomain
$i+1$ and go back to step 2.

\item Otherwise, a new superspin $(i+1)'$ is created, a distance
$\del$ to the right of the old one which is distributed according to
%
%$P(\del)\propto\exp\{-[\del-(\ddr-\dd_i)]\} / [1-\exp(-\dd_i)]
%
$P(\del)\propto\exp(-\del)$ over the range
$\del\in[\ddr-\dd_i,\ddr]$. This gives new superdomain
lengths: $\dd_{i}\leftarrow \dd_{i}+\del$ and $\dd_{i+1}\leftarrow
\dd_{i+1}-\del$. If $\dd_{i+1}\leq\ddr$, increase $i\leftarrow i+1$
and go back to step 3. Otherwise the relaxation process at this $\ddr$
is complete; go back to step 2.

\end{enumerate}

Ideally, one would like to solve the above superdomain model directly
and thus compute the dependence of the persistence function on $\ddr$
analytically. The iterative process of superspin elimination and
regeneration is difficult to keep track of, however, and we have not
been able to find an analytical solution. Nevertheless, because the
superdomain model expresses times in terms of the lengthscales $\ddr$,
which unlike the timescales themselves do not diverge for $T\to 0$, it
can be simulated easily and accurately; we have used systems of $10^5$
superdomains (at $\ddr=0$), checking that finite-size effects are
negligible and typically averaging our results over $10^4$ simulation
runs.
\begin{figure}
\epsfig{file=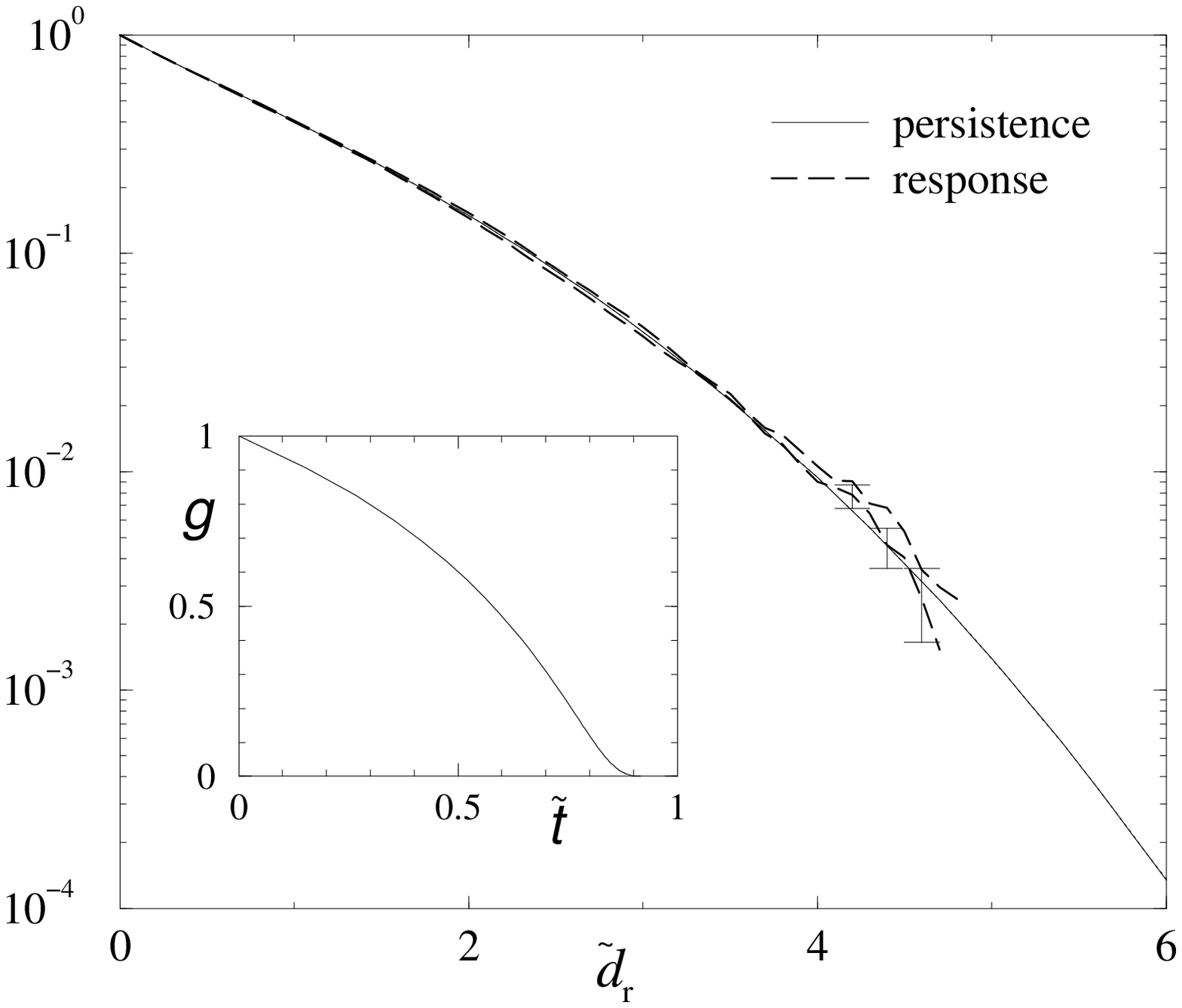,width=8.4cm}
\caption{Predictions of the superdomain model. Main plot, dashed
lines: Response functions for $\eps_0/\eps=0.95$ and
$\eps_0/\eps=1.05$ versus relaxation `time' $\ddr$; on the latter, a
few error bars are shown where they are significant.  Solid lines:
Persistence functions $\pup$ and $\pdown$; the two lines are
indistinguishable by eye. The inset shows the scaling function
$g(\dt)$ for the equilibrium relaxation functions, as derived from the
superdomain persistence and the scaling function for the FPTs sketched
in Fig.~\protect\ref{fig:scaled_MFPTs}(left).
%
%$\pdown/\pup-1$ (solid) and
%the relative errors of $\pup$ and $\pdown$, demonstrating that
%$\pdown=\pup$ to within numerical accuracy.
%
\label{fig:persistence}
}
\end{figure}

The key quantity that we want to predict from the superdomain model is
the persistence function. The up-spin persistence $\pup$ is the
fraction of up-spins that have never flipped since $\ddr=0$. Since at
$\ddr=0$ all up-spins are superspins, $\pup$ is the fraction of
these initial superspins that have never been equilibrated. We show
the results in Fig.~\ref{fig:persistence}; $\pup$ initially decreases
linearly with $\ddr$, but then the decay becomes much faster and
indeed super-exponential in $\ddr$.

If, as we have claimed, the superdomain model is the correct effective
description for the $T\to 0$ dynamics of the East model, then it must
obey the exact identity $\pup=\pdown$ derived in
Sec.~\ref{sec:functions}. $\pdown$ is measured in the superdomain
model as the fraction of the chain which has never been swept by an
equilibrated zone, a quantity that one might not have naively
suspected to be connected to the number of persistent
superspins. Nevertheless, our simulations show that indeed
$\pup=\pdown$ to very high accuracy ($\sim 1\%$, less than the
relative error on the measurements of $\pup$ and $\pdown$) in the
superdomain model. In fact, both quantities are plotted in
Fig.~\ref{fig:persistence} but are indistinguishable by eye. This
provides strong support for the correctness of the superdomain model;
if \eg\ the mechanism for regenerating superdomains is neglected,
one finds that the condition $\pdown=\pup$ is violated.

A further consistency check on the model is obtained from the
requirement that it should represent the {\em equilibrium}
dynamics. The concentration of up-spins in the system, \ie\ the
magnetization $M=(1/L)\sum_i s_i$, should thus remain independent of
$\ddr$. Within the superdomain model, if there are $\Ns$ superspins at
time $\ddr$, one has
\be
M=\Ns(1+c\ddr/\eps)/L \approx (\Ns/L)(1+\ddr)
\label{M}
\ee
since in addition to the superspins there are a further $c\times
\Ns\ddr/\eps$ up-spins on the $\Ns\ddr/\eps$ sites covered by the
equilibrated zones. This quantity should equal $M=c\approx \eps$
independently of $\ddr$. Hence the average (rescaled) distance between
superspins, $\eps L/\Ns$, which is identical to the average superdomain
length, must equal $1+\ddr$ at time $\ddr$. This is indeed what our
simulations of the superdomain model show. One can push this
comparison further and consider not just the total up-spin
concentration, but the distribution of sizes of the domains formed by
all spins (rather than just the superspins). This distribution must be
a pure exponential, independently of $\ddr$. It can be expressed in
terms of the superdomain distribution by an appropriate convolution
which accounts for the fact that additional up-spins exist in the
equilibrated zones. Omitting the details, we only state that one finds
in this way that the superdomain size distribution at time $\ddr$ must
be
\be
P(\dd;\ddr) = \Theta(\dd-\ddr) {\rm e}^{-(\dd-\ddr)}
\label{superdomain_dist}
\ee
Again we find that this is verified in our simulations. Note
that\eq{superdomain_dist} has a simple intuitive interpretation: it
corresponds to an exponential distribution of the segments of
down-spins separating the equilibrated zone of each superdomain from
the next superspin on the right.

A final check on the superdomain model is that it should obey
FDT---since the original model obeys detailed balance FDT is
automatically satisfied but this is not guaranteed for the superdomain
description. As explained in Sec.~\ref{sec:functions}, all non-local
correlation (and hence response) functions vanish in the East model,
so that one is free to consider either a local response of $s_i$ to a
local field, or a response of the magnetization $M$ to a uniform field
$h$. Choosing the second option, the energy function is modified to $E
= (1-h)\sum_{i} s_i$ which is equivalent to changing
$\eps$ to $\eps'=\eps\exp(h/T)$ (or temperature from $T$ to
$T/(1-h)$). The response to a field $h$
switched on in the distant past and switched off at $t=0$ can thus be
measured by initializing the system in an equilibrium state
corresponding to $\eps'$ and monitoring the evolution of $M$ during
the subsequent dynamics at $\eps$. By FDT this switch-off response
should have the same time-dependence as the correlation (and hence the
persistence) functions. In the superdomain model, the measurement is
performed by initializing the superdomains with a modified domain size
distribution $P(\dd)=(\eps'/\eps)\exp[-(\eps'/\eps)\dd]$ and then
tracking the decay of the magnetization, measured as in\eq{M},
to its equilibrium value. The response functions simulated for
$\eps'/\eps=0.95$ and $1.05$ are plotted in Fig.~\ref{fig:persistence}
above, and show that the superdomain model indeed obeys FDT.

Pushing the above scenario further, one could consider nonlinear
responses in the superdomain model, in particular a large ratio
$\eps'/\eps$ which corresponds to a quench to a much lower
temperature. (For the superdomain model to remain applicable we still
need $\eps'\ll 1$, of course.) The initial scaled domain lengths in
the system
are then of order $\eps/\eps' \ll 1$. In the regime of small $\ddr$
where these domain lengths are removed from the system, one sees that
the probability $1-\exp(-\ddr)\approx \ddr$ of creating a new superspin
is very small, tending to zero for $\ddr\sim\eps/\eps'\to 0$. In
this limit the superdomain dynamics becomes `take the smallest
superdomain and coalesce it with its neighbour on the right', which
is precisely the paste-all model discussed in
Sec.~\ref{sec:paste-all}.  While superdomain sizes remain $\ll 1$, the
domain size distribution will thus be driven to the scaling
distribution of the paste-all model. As demonstrated in
Sec.~\ref{sec:paste-all}, this scaling distribution is the {\em same}
as that for the coarsening dynamics of the East model. Thus, the
prediction of the superdomain model for the form of the domain size
distribution, a long time after a quench from $\eps'$ to $\eps$ with
$1\gg\eps'\gg \eps$, matches up precisely with that predicted in
Sec.~\ref{sec:solution} for a quench from $1\approx
\eps'\gg\eps$. (Note that in the first case we are in principle
talking about the {\em superdomain} size distribution, not the actual
domain distribution as in Sec.~\ref{sec:solution}. However, for small
$\ddr$ the two are identical since the number of up-spins within the
equilibrated zones of the superdomains are negligible.)

Writing our result for the persistence function in the superdomain
model (either $\pup$ or $\pdown$, since they are identical) as
$\psd(\ddr)$, we can now translate it into a scaling prediction for
the time-dependence of the correlation and persistence functions at
low $T$ in the East model. Using the inverse of the scaling function
$\scf(\dd)$ from\eq{timescaling}, one has
\be
C(t)=\pup(t)=\pdown(t)=\psd\left( \scf^{-1}(\dt) \right), \quad
\dt = \left(\frac{t}{\tst}\right)^{T \ln 2}
\label{final_scaling}
\ee
Thus the relaxation functions show strong stretching, with the
stretching exponent $T\ln 2$ decreasing to zero for $T\to 0$. However,
the scaling function $g(\dt)=\psd(\scf^{-1}(\dt))$ is
non-exponential. In fact, since $\scf(\dd)$ approaches a constant
$\scf_\infty$ for $\dd\to\infty$ as argued after\eq{timescaling}, the
scaling function in\eq{final_scaling} decays to zero at a {\em finite}
value $\dt=\scf_\infty$. On the basis of the results of\cite{AldDia02}
we would conjecture this value to be $\scf_\infty=1$, see
after\eq{timescaling}; the scaling function $g(\dt)$ for this case is
sketched in the inset of Fig.~\ref{fig:persistence}. It is important
to bear in mind that the limit of $T\to 0$ considered here is taken at
constant $\ddr$ or, equivalently, constant $\dt$. If we look instead
at fixed nonzero $T$ then the relaxation functions are of course
nonzero for all $t$, but this does not contradict $g(\dt)$ dropping to
zero at $\dt=\scf_\infty$. To see this, note that the asymptotic decay
of all relaxation functions should be to leading order an exponential
with the longest relaxation time, $\sim \exp(-t/\tst)$. This does
remain nonzero for all $t$; but in terms of the scaled time $\dt$ it
becomes $\sim\exp(-\dt^{1/T\ln 2})$ which converges to zero for $T\to
0$ at any fixed $\dt>1$.

We can now also come back to a point discussed at the end of
Sec~\ref{sec:FPT}: in principle the function $\scf(\dd)$ may not be
strictly monotonically increasing, but could instead be {\em constant}
from a certain value $\dd_*$ onwards, $\scf(\dd)=\scf_\infty$ for
$\dd>\dd_*$. Looking at\eq{final_scaling}, however, this would lead to
the conclusion that for any $\dt<\scf_\infty$ the relaxation functions
are nonzero, approaching a nonzero limit as $\dt\to\scf_\infty$ but
then dropping discontinuously to zero (in the limit $T\to 0$). This
makes this hypothetical behaviour of $\scf(\dd)$ rather unlikely.

\begin{figure}
\epsfig{file=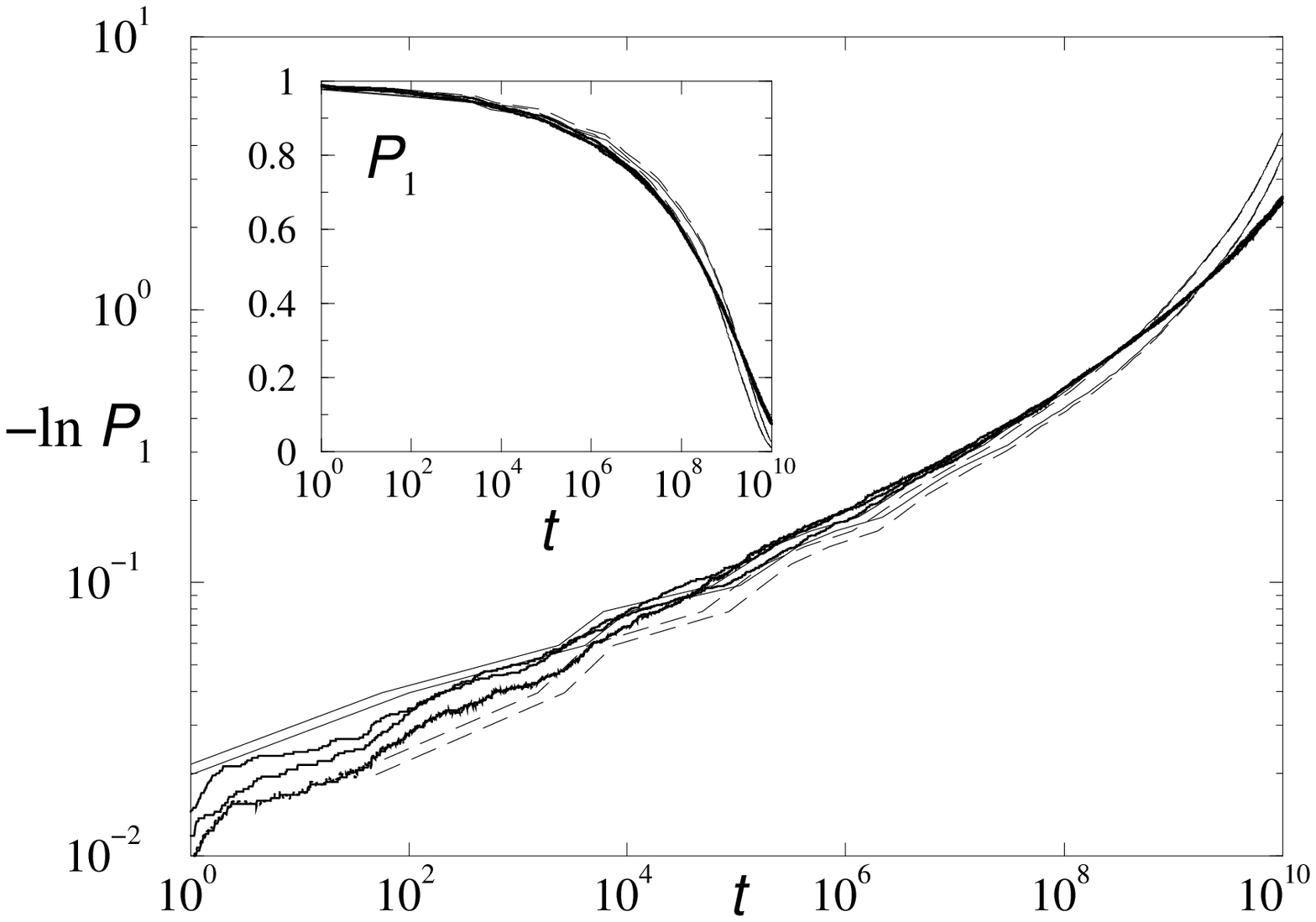,width=8.4cm}
\caption{Comparison of up-spin persistence function $\pup$ from
simulations with prediction of superdomain model, for $\eps=0.02$. A
log-log plot of $-\ln \pup(t)$ is shown, which would be straight for
stretched exponential relaxation. The inset shows $\pup(t)$ directly.
Thick lines: $\pup$ as simulated in three runs for a chain of length
$L=2^{17}$; the correlation function $C(t)$ is also shown for one run
but is indistinguishable from $\pup(t)$ as predicted for low
$\eps$. Thin lines: Predictions of the superdomain model (see
text). Dashed/solid: $t$ determined from FPTs for up- and down-flips,
respectively; left/right curve of each pair: using mean of the log-FPT
and the log of the MFPT, respectively.
\label{fig:comparison}
}\end{figure}
As explained,
one of the main benefits of the superdomain model is that it allows
one to access very long timescales that diverge extremely quickly as
$T$ decreases. Since this is precisely the regime that is hard to
probe with simulations, comparing the theoretical predictions with
numerical data is not straightforward. We chose $\eps=0.02$, the
lowest value for which we can simulate a significant part of the
equilibrium relaxation. As Fig.~\ref{fig:scaled_MFPTs} shows, the
relation between rescaled FPTs $\dtau$ and rescaled distances
$\dd=\eps d$ has not yet reached its $\eps\to 0$ form, so it would
make no sense to compare numerical data with superdomain predictions
based on the latter. Instead, we use the numerically obtained FPTs to
link timescales and lengthscales. In detail, we obtain for each
unnormalized distance $d$ the measured FPT, and plot against this time
the superdomain model's persistence function $\psd(\dd)$ at the
relevant scaled distance $\dd=\eps d$. Since we measured FPT in four
different ways---for up- and down-flips, and averaging FPTs or
log-FPTs---which we expect to become identical only for $T\to 0$, this
procedure gives four slightly different curves for the superdomain
predictions. As shown in Fig.~\ref{fig:comparison}, these bracket the
numerically simulated up-spin persistence and correlation functions
rather well, especially given that there are no fit parameters in the
comparison. The largest deviations occur for large times,
corresponding to large lengthscales. This is consistent with
Fig.~\ref{fig:scaled_MFPTs}, which shows that in this regime the
behaviour still differs substantially from that expected in the limit
$T\to 0$.

\section{Conclusion}
\label{sec:conclusion}

%Future work: Maybe study dynamical heterogeneities using superdomain
%model. Mention confusion over this point?

We have studied the dynamics of the East model. Consisting of 
uncoupled spins in a downward-pointing field, the model has trivial
equilibrium statistics. However, the kinetic constraint that spins can only
flip if their left neighbour is up causes pronounced glassy features
in the dynamics at low $T$, when the concentration of up-spins is low.

We first studied the non-equilibrium coarsening dynamics after a
quench to low $T$. In the limit $\eps=\exp(-1/T)\to 0$ the equilibrium
concentration of up-spins at the new temperature is negligible
($\approx\eps$) and the flipping-down of up-spins becomes irreversible
to leading order. This allows the dynamics to be described as
coarsening via coalescence of down-spin domains. The process is
hierarchical, being governed by a series of well-separated
timescales. We solved this hierarchical coarsening dynamics exactly,
using an independent intervals method that becomes exact for $T\to
0$. Anomalous coarsening results, with typical domain lengths scaling
as $\dbar \sim t^{T \ln 2}$. The dominant divergence of the
equilibration time for low $T$ can also be estimated, and is given by
the factor $\tst=\exp(1/T^2\ln 2)$, an EITS dependence typical of
fragile glasses. For large domain sizes $\dbar$ that are still small
compared to the equilibrium value $\deq$, the domain size distribution
approaches a scaling form. We showed that this scaling distribution is
equal to that of the paste-all model, and were able to define a whole
family of interpolating models that all share this scaling
distribution.

In the second part of the paper we focussed on the equilibrium
dynamics at low $T$. We showed that the standard relaxation functions,
spin-autocorrelation and persistence of up- and down-spins, become
identical for $T\to 0$, so that only one of them needs to be
considered. We then investigated the relation between time- and
lengthscales. Generalizing from the results in the coarsening regime
$\dd=d/\deq\ll 1$, we introduced a time scaling hypothesis. This
implied that for $T\to 0$ one has continuous timescale separation,
with domains of any two different scaled sizes $\dd$ relaxing on
well-separated timescales. On this basis we proposed a model of
superdomains, which are bounded by up-spins that are frozen on long
timescales. The dynamics of this model is nontrivial and not, as yet,
analytically tractable, but can easily be simulated since time is
effectively measured in terms of the scaled distance $\dd$, whose
relevant values remain $\order(1)$ even for $T\to 0$. We verified that
the model obeys important consistency requirements, in particular the
equality of up-spin and down-spin persistence, the
fluctuation-dissipation theorem, and the stationarity of up-spin
concentration and domain size distribution in equilibrium.

From the superdomain model predictions we finally deduced that the
equilibrium relaxation functions should decay for low $T$ as $g(\dt)$,
with $g(\cdot)$ a scaling function of $\dt=(t/\tst)^{T\ln 2}$. This
demonstrates strong stretching for low $T$, but the overall relaxation
is more complicated than a stretched exponential. In fact, the
function $g(\cdot)$ decays faster than exponential, and in the limit
$T\to 0$ at fixed $\dt$ reaches zero at a {\em finite} value of
$\dt$. The lowest temperature $T$ that we can conveniently simulate,
corresponding to $\eps\to 0.02$, is still rather far from the
asymptotic $T\to 0$ limit but nevertheless showed reasonable agreement
between numerical simulations and appropriately extracted predictions
of the superdomain model. We would suggest that stretching but not
simple stretched {\em exponential} behaviour may be rather generic in glassy
dynamics. Actual stretched exponentials could more often than
not be just convenient fitting functions over a limited number of
decades in time. To clarify this point, a study of low-temperature
relaxation in other solvable models exhibiting glassy dynamics would
obviously be desirable.

In future work, it would be interesting to see whether the
out-of-equilibrium response of spins to a local field could
also be analysed within the irreversible coarsening framework we used
above. This response function was simulated
in~\cite{CriRitRocSel00} and found there to be monotonic; at lower
$T$, however, non-monotonicities should appear according to a later
conjecture~\cite{GarNew00}. A closer investigation of
out-of-equilibrium FDT relations would also be worthwhile. Previous
results for these~\cite{CriRitRocSel00} have to be regarded with some
caution since they were constructed using a disconnected correlator;
see \eg~\cite{SolFieMay02} for discussion of this point. Finally, as
regards the equilibrium dynamics, it will be interesting to analyse
the implications of the superdomain description for dynamic
heterogeneities, making a connection to the recent work of Garrahan
and Chandler~\cite{GarCha02}.

%%%%%%%%%%%%%%%%%%%%%%%%%%%%%%%%%%%%%%%%%%%%%%%%%%%%%%%%%%%%%%%%%%%%%%%%%%%%%%

\begin{acknowledgments}
PS acknowledges
financial support through Nuffield grant NAL/00361/G.
\end{acknowledgments}

\bibliographystyle{prsty}
\bibliography{/home/psollich/references/references,local}

\begin{thebibliography}{10}
\bibitem{irrev}
What we mean by irreversibility to leading order is that, although down-spins
  do flip back up during the relaxation processes that cause the domain
  coarsening, they have a negligibly small probability of being up in a
  snapshot at some randomly chosen time.
\end{thebibliography}

\end{document}